\documentclass[aps,prb,preprint]{revtex4-1}
\usepackage{amsmath}
\usepackage{graphicx}
\usepackage{amssymb}

\usepackage{color} 
\usepackage{graphicx}
\usepackage{bm}
\usepackage{amsmath}

\usepackage{gt17}

\DeclareMathAlphabet\mathbfcal{OMS}{cmsy}{b}{n}

\begin{document}
\title{
Spin correlation function theory of spin-charge conversion effects
}
\author{Gen Tatara}
\affiliation{RIKEN Center for Emergent Matter Science (CEMS)
and RIKEN Cluster for Pioneering Research (CPR), 
2-1 Hirosawa, Wako, Saitama, 351-0198 Japan}

\date{\today}

\begin{abstract}
Theory of spin-charge conversion effects in spintronics are presented in terms of correlation functions of physical observables, spin and electric current. 
Direct and inverse spin Hall effects and spin pumping effect are studied considering metallic systems with random spin-orbit interaction and spatially nonuniform Rashba interaction.  
The theory is free from ambiguity associated with spin current, and provides a clear physical picture of the spin-charge conversion effects. In the present approach, the spin current transmission efficiency turns out essentially to be the nonuniform component of magnetic susceptibility.
\end{abstract}  

\maketitle
\newcommand{\correlation}{C}

\section{Introduction}
Spintronics phenomena, transport and mutual conversion of spin and charge in solids, are conventionally explained in terms of flow of spin, spin current, by analogy with the case of electric current. 
Physically, however, there is a fundamental difference between transports of electric charge and spin in solids, namely, the former is conserved, while the latter is not.

The electric charge density $\rho$ and current $\jv$ satisfy the conservation law, 
$\dot{\rho}+\nabla\cdot\jv=0$.
Steady configuration of charge distribution is realized when divergence of current is zero, $\nabla\cdot\jv=0$. 
Because of the conservation law, the amount of electric charge accumulated is counted by estimating the total current. 
Therefore the electric conductivity, which is expressed by current correlation function, is directly related to the electric permittivity, expressed by a correlation function of  electron density.  
In other words, electric transport properties can be described either by use of scalar or vector potentials, owing to the gauge invariance.

In contrast, spin current, $j_{{\rm s},i}^\alpha$ ($\alpha$ and $i$ denote direction of spin and flow, respectively), is not conserved but satisfies a continuity equation,
\begin{align}
 \dot{s}_\alpha+\nabla\cdot \jv_{\rm s}^\alpha={\cal {T}}_\alpha,\label{spincontinuity}
\end{align}
where $\sv$ and  $\mathbfcal{T}$ denote spin density and spin relaxation torque, respectively, and the divergence is with respect to the spatial direction of flow.
The relaxation torque is phenomenologically expressed as proportional to the induced nonequilibrium  spin density as
\begin{align}
\mathbfcal{T} =\frac{\sv}{\tau_{\rm r}},\label{relaxationtorque}
\end{align}
where $\tau_{\rm r}$ is a relaxation time of the process concerned. 
The steady nonequilibrium spin configuration is therefore determined by the balance of flow and relaxation of spin as  
\begin{align}
s_\alpha = \tau_{\rm r} \nabla\cdot\jv_{\rm s}^\alpha . \label{spinequilibrium}
\end{align}
The (divergence of) spin current is thus detectable by measuring nonequilibrium spin accumulation.
One must note, however, that definition of spin current is not unique because of its nonconvservation. 
Conversely, relaxation time $\tau_{\rm r}$ depends on the definition of spin current.
These ambiguities are crucial  both physically and quantitatively for interpreting spin transport experiments in terms of spin current on a phenomenological basis.

The objective of the present paper is to describe spin-charge conversion phenomena in terms of physical observables without referring to spin current.
We do this by explicit calculation, considering simple models of weak spin-orbit interaction arising from either impurities or localized Rashba interaction.
The spin Hall (SH) effect is described by directly calculating spin accumulation induced by applied electric field by evaluating correlation function of spin and electric current, $C_{SJ}$.
The idea is similar to the original argument of spin Hall effect by Dyakonov \cite{Dyakonov71}.
The result is shown to be consistent with conventional spin current interpretation \cite{Hirsch99} and experiments\cite{Kato04}.
The inverse spin Hall effect is also discussed, considering two cases of 'spin current injection', one by spin pumping  effect and the other by spin Hall effect.
The inverse spin Hall effect corresponds to correlation of electric current and spin, $C_{JS}$, the reciprocal of $C_{SJ}$ for spin Hall effect. 
For spin pumping we consider the case of metallic ferromagnet. 
Physically, spin pumping effect is driven by non-equilibrium spin gauge field, which generates nonequilibrium spin accumulation at the interface \cite{TataraSP17}, and which may be called the non-adiabatic spin chemical potential. 
A spin current then arises by electron diffusion and is proportional to the gradient of spin density, consistent with the picture originally presented by Silsbee \cite{Silsbee79}.
The expression of generated 'spin current' agrees with conventional spin pumping formula presented by Tserkovnyak et. al. \cite{Tserkovnyak02}. 
It is essential to note that the spin pumping generates interface spin accumulation, and not spin current.
The correct driving field for the inverse spin Hall effect is therefore the interface spin accumulation, and the corresponding physical correlation function is $C_{JS}$.

For combination of spin Hall and the inverse spin Hall effects, we consider junction of a nonmagnetic spacer and two heavy metal contacts for measurement and external electric field, the setup called non-local spin injection.
The magnitude of the output current is shown to be represented by a product of correlation functions of charge current and spin, $C_{JS}C_{SJ}$, of heavy metal and spin correlation function $C_{{\rm N}SS}$ of nonmagnetic metal (N).
The transmittancy of  spin current in normal metal is represented by spin correlation function or magnetic susceptibility in the present scheme.

Those spin-charge conversion and spin transport phenomena turn out to be described elegantly without ambiguity in terms of correlation function of physical observables, spin and electric current. 
The demonstration here is carried out, however, on simple theoretical models, and we do not claim generality. 
Nevertheless, the present formulation has potential of wide applicability. For instance, including interaction effects with magnons and phonons in the correlation functions or to consider insulators or antiferromagnets are straightforward. 
Quantitative predictions shall be given by numerically calculating the Green's functions on realistic tight-binding models.

Spin transport has been discussed in a number of theoretical works \cite{Burkov04,Galitski06,Tokatly10,Shibata11}.
Most studies are devoted to deriving the kinetic equation (diffusion equation), equivalent to the continuity equation, for non-equilibrium spin density.
For discussing spin transport based on the kinetic equation, boundary condition plays crucially important role, as was pointed out in Ref. \cite{Galitski06}.
The motivation of our approach is different from those based on the kinetic equation; We calculate the induced spin density directly by use of a linear response theory instead of solving the kinetic equation. 
In the case of spin Hall effect, the induced spin density when uniform electric field is applied is obviously  not spatially uniform.
We need therefore to look into the non-uniform component of the response function, namely at finite wave vector $q$ of the external field.
This is in contrast to the conventional formulation in terms of spin current. In fact, the response function of spin current and electric current has a finite uniform component, resulting in a uniform spin Hall conductivity. This description seems so far convenient, although physical observable, spin density, is obtained only after  by solving the diffusion equation.
In contrast, what is proposed in the present paper is to calculate physical observable within a single framework of linear response theory by considering non-uniform ($q\neq0$) component of the response function.

In the context of current-induced torques in ferromagnets, the present approach is straightforward and natural, as the torque is calculated by evaluating non-equilibrium spin density \cite{TK04,TKS_PR08,YuanSOT16}.
In fact, spin-orbit torque in a bilayer was recently studied  avoiding notion of spin current in Ref. \cite{Ado17}.

\section{Spin-charge conversion due to impurity spin-orbit interaction}

Let us start microscopic calculations of spin-charge conversion effects. In this section we consider the case of spin-orbit interaction induced by random impurities,  represented by a Hamiltonian  
\begin{align}
 H_{\rm so(i)} &=   \lambda  \intr c^\dagger[(\nabla \Vi(\rv)\times {\pv})\cdot \sigmav] c,
 \label{Hsoi}
\end{align}
where $\pv$ is electron momentum, $\lambda$ is the strength of the spin-orbit interaction, and $\Vi(\rv)$ is the impurity potential, which we treat as point-like, i.e., $\Vi(\rv)=\Vi\delta(\rv-\Rv_i)$, where $\Vi$ is the strength and $\Rv_i$ is random impurity position.
$\hbar$ is set to unity.
We define current-spin correlation function, $\correlation_{JS}$, which represents the conversion efficiency of spin density to charge current, as 
\begin{align}
 \correlation_{JS}^{\alpha\beta}(\qv)\equiv \sum_{\kv}\tr[{v}_\alpha G^\ret_{\kv+\qv} \sigma_\beta G^\adv_{\kv}], \label{correlationdef}
\end{align}
where ${\vv}$ is the velocity operator and $G^\ret$ and $G^\adv$ are retarded and advanced Green's functions at zero angular frequency, respectively, including interactions such as spin-orbit and impurity scatterings, and $\alpha, \beta=x,y,z$ denote index for space and spin.  
This definition of the correlation function is focusing on the dominant contribution in the limit of vanishing external frequency of the full correlation function,   
 $\chi_{JS}^{\alpha\beta}(\qv,\Omega)\equiv \sumom\sum_{\kv}\tr[{v}_\alpha G_{\kv+\qv}(\omega+\Omega) \sigma_\beta G_{\kv}(\omega)]^<$, where $\omega$ and $\Omega$ are angular frequencies of electron and external source (see Sec. \ref{SEC:correlation}).

 \begin{figure}
 \includegraphics[width=0.2\hsize] {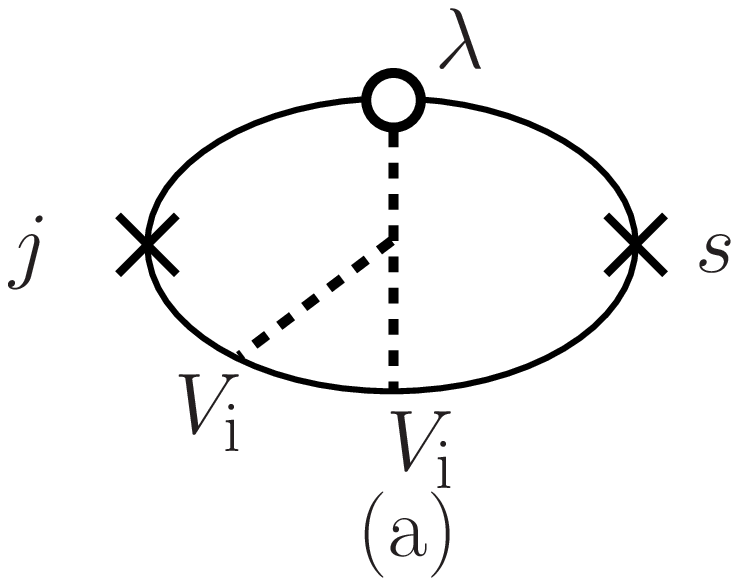}
 \includegraphics[width=0.2\hsize] {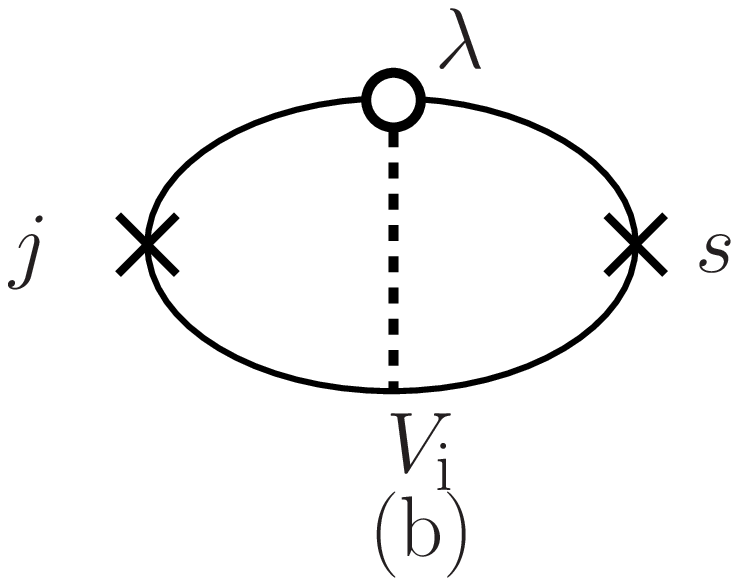}
 \includegraphics[width=0.2\hsize] {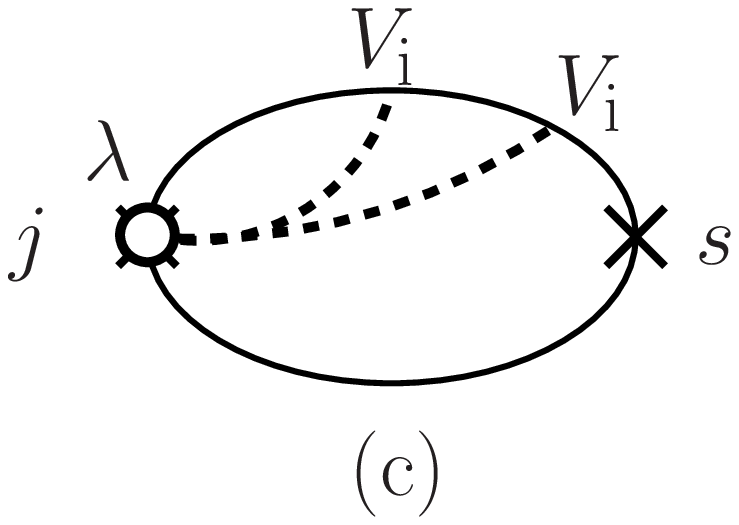}
\caption{ Diagramatic representation of current-spin correlation function including spin-orbit interaction due to random impurity to the linear order. 
Diagram (a) is dominant contribution, while 
contributions (b) and (c) vanish or small. 
\label{FIGSCC}}
 \end{figure}
\begin{figure}
 \includegraphics[width=0.4\hsize] {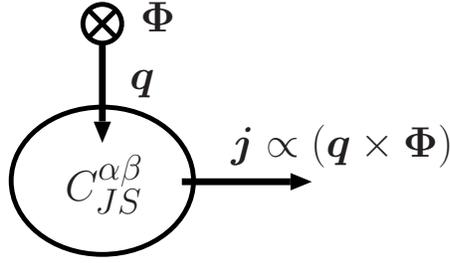}
\caption{ Schematic figure showing spin-charge conversion property of $C_{JS}^{\alpha\beta}(\qv)\propto i \epsilon_{\alpha \beta\gamma} q_\gamma$. 
When a spin source field, $\bm{\Phi}$, changing in the direction of $\qv$ ('spin flow' direction), is applied, electric current $\jv$ is generated in the direction perpendicular to  $\bm{\Phi}$ and $\qv$ ($j_\alpha\propto \epsilon_{\alpha \beta\gamma} q_\gamma \Phi_\beta$). This is in agreement with conventional spin-charge conversion formula, $j_\alpha\propto \epsilon_{\alpha \beta\gamma} j_{{\rm s},\gamma}^\beta$, as $j_{{\rm s},\gamma}^\beta$ corresponds to $q_\gamma \Phi_\beta$ in the present formalism. 
\label{FIGCJS}}
 \end{figure}

 We consider the lowest order (first order) contribution of the spin-orbit interaction. 
The dominant contribution describing spin-charge conversion is described by the diagram of Fig. \ref{FIGSCC}(a), where the momentum conservation is recovered after averaging taking account of the second-order scattering by normal impurities.
Noting that electrons are not spin-polaized, it reads (including the complex conjugate process) 
\begin{align}
 \correlation_{JS}^{\alpha\beta}(\qv)= 2i \frac{\lambda \nimp\Vi^3}{m}\sum_{\kv\kv'\kv''} \epsilon_{ij\beta}k_\alpha
 g^\ret_{\kv+\frac{q}{2}}g^\adv_{\kv-\frac{q}{2}} g^\ret_{\kv'+\frac{q}{2}}g^\adv_{\kv'-\frac{q}{2}} 
 \lt[ 
 \lt(k'+\frac{q}{2}\rt)_i\lt(k+\frac{q}{2}\rt)_j  g^\adv_{\kv''} 
-\lt(k'-\frac{q}{2}\rt)_i\lt(k-\frac{q}{2}\rt)_j g^\ret_{\kv''} 
\rt].\label{chijsdef}
\end{align}
The contribution including normal impurities to the linear order (Fig. \ref{FIGSCC}(b)) vanish, similarly to the case of anomalous Hall effect \cite{Crepieux01}.
Other processes like Fig. \ref{FIGSCC}(c) are smaller by order of $(\ef\tau)^{-1}$, where $\tau$ is elastic lifetime of electron. 
We consider the case of free electron Green's function, $ g^\ret_{\kv}=[-\ekv+\frac{i}{2\tau}]^{-1}$ with quadratic dispersion, $\ekv\equiv \frac{k^2}{2m}-\ef$, $\ef$ being the Fermi energy. 
Elastic inverse lifetime $\tau^{-1}=\tau_0^{-1}+\tau_{\rm so}^{-1}$ has two contributions, $\tau_0^{-1}$  from normal impurity scattering and $\tau_{\rm so}^{-1}$ from spin-orbit interaction. The normal impurity contribution satisfies $2\pi\dos\nimp\Vi^2\tau_0=1$, where $\dos$ is the electron  density of states at the Fermi energy, and $n_{\rm i}$ is impurity concentration. 
The difference between $\tau$ and $\tau_0$ is negligible in the ballistic consideration in this section, but is essential in discussing diffusive contribution.
Using $ \sum_{\kv}g^\adv_{\kv}=i\pi\dos$ and 
\begin{align}
\sum_{\kv} k_\alpha \lt(k+\frac{q}{2}\rt)_j g^\ret_{\kv+\frac{q}{2}}g^\adv_{\kv-\frac{q}{2}} &\simeq \delta_{\alpha,j} \frac{\kf^2}{3}2\pi\dos\tau \label{sumk1} \\
\sum_{\kv'} \lt(k'+\frac{q}{2}\rt)_i g^\ret_{\kv'+\frac{q}{2}}g^\adv_{\kv'-\frac{q}{2}} &\simeq -i \frac{\kf^2}{3m}2\pi\dos\tau^2 q_i, \label{sumk2}
\end{align}
we have 
\begin{align}
 \correlation_{JS}^{\alpha\beta}(\qv)= i\lambda_{\rm sh} \epsilon_{\alpha\beta i} q_i,
 \label{chijsresult}
\end{align}
where
\begin{align}
\lambda_{\rm sh}\equiv \frac{2}{3}(2\pi)^2 \epsilon_{\rm so} \dos^2 D\tau,
\end{align}
$D\equiv \frac{\kf^2\tau}{3m^2}$ is electron diffusion length and $\epsilon_{\rm so}\equiv \lambda \Vi \kf^2$ is the energy scale of spin-orbit interaction.
For the slowly-varying case we consider, the correlation function $\correlation_{JS}^{\alpha\beta}(\rv)$ is local and is proportional to linear spatial derivative. 
Equation (\ref{chijsresult}) clearly indicates that 'spin-charge conversion' mechanism works on the gradient of spin density, in the same manner as on the spin current. 
The direction of flow of spin, $i$ in Eq. (\ref{chijsresult}), is perpendicular to both spin ($\beta$) and electric current direction ($\alpha$), in agreement with phenomenological spin-charge conversion picture \cite{Saitoh06}, postulating $j_{{\rm s},i}^\beta\propto \epsilon_{i\alpha\beta}j_\alpha$ and 
$j_\alpha\propto \epsilon_{i\alpha\beta}j_{{\rm s},i}^\beta$. 
Geometry of spin-charge conversion property of $C_{JS}^{\alpha\beta}$ is shown in Fig. \ref{FIGCJS}.
In the next section, we show that the result, spin density is induced by a spatial derivative of the applied electric field, reproduces conventional spin Hall effect.

%

\section{Spin Hall effect}

Based on the current-spin correlation function, Eq. (\ref{chijsresult}), we discuss the spin Hall effect.  
The spin density induced by applying uniform electric field $\Ev$ is given by 
\begin{align}
 s_\alpha= i \correlation_{SJ}^{\alpha\beta} E_\beta,
 \end{align}
where we represented the electric field in terms of a vector potential as $\Ev=-\dot{\Av}$ and evaluated the linear response contribution (See Sec. \ref{SEC:correlation}). 
 Namely, the spin Hall formula we have is 
\begin{align} 
\sv= \lambda_{\rm sh}(\nabla \times  \Ev)=\frac{\lambda_{\rm sh}}{\sigma_{\rm e}}(\nabla \times  \jv),\label{SHs}
\end{align}
where $\sigma_{\rm e}$ is electric conductivity.
This simple equation indicates that spin Hall effect induces an inhomogeneous spin accumulation when an electric field is applied. Assuming homogeneous applied current in the bulk, spin accumulation is formed at the edge of the system, where the applied current vanishes.
This result is consistent with experimental observation and previous theories indicating importance of boundary \cite{Galitski06}. 
The actual spin profile shall be discussed taking account of electron diffusion in the next section.

Equation (\ref{SHs}) indicating inhomogeneous spin accumulation is consistent with conventional spin current picture discussed on phenomenological grounds, as spin current and gradient of spin are naively the same. 
The equivalence is confirmed microscopically by calculating the correlation function of electric current and  spin current vertex, which we define without the spin-orbit interaction as 
$\frac{k_\beta}{m}\sigma_\alpha$.
The current-spin current correlation function then is obtained by replacing $-i\tau q_i$ in Eq. (\ref{sumk2}) by $\delta_{i\beta}$, where $\beta$ is the direction of spin current flow. This leads to a spin Hall formula in more conventional form, i.e.,
\begin{align}
 j_{{\rm s},\beta}^\alpha =\frac{\lambda_{\rm sh}}{\tau} \epsilon_{\alpha\beta\gamma}E_\gamma .
 \label{SHjs}
\end{align}
Conventional argument use the expression for the spin current to discuss spin accumulation generated at the edges, using spin continuity equation, Eq. (\ref{spincontinuity}).
The result (\ref{SHs}) is consistent with the conventional picture using Eq. (\ref{SHjs}) and (\ref{spincontinuity}), as it was obtained in the ballistic regime (or for shorter time scale compared to $\tau$), where the electron elastic lifetime is the only relevant relaxation time and thus spin relaxation time coincides with the elastic one, $\tau_{\rm r}=\tau$.
In other words, momentum scattering at short time scale leads effectively to a spin relaxation, as it modifies electron state and affects nonequilibrium spin accumulation. 
In diffusive regime, the relaxation time is replaced by a longer time scale of electron spin lifetime, as we shall see in the next section. 

Solving a diffusion equation for electron spin in the presence of spin current in the conventional spin current analysis corresponds mathematically to deriving spin density by evaluating $\nabla\cdot\jv_{\rm s}$ and spin relaxation torque $\mathbfcal{T}$.  In this conventional approach, however, ambiguities of definition of  spin current and spin relaxation torque arise if carried out phenomenologically. 
(These ambiguities are not physical; In fact there is no ambiguity if carried out fully microscopically as the form of the relaxation torque is uniquely determined by the definition of spin current \cite{TE08,Nakabayashi10}.)
Introducing such ambiguity for an indirect explanation of observables using the concept of spin current seems physically awkward,    
in the viewpoints of Eq. (\ref{SHs}), which gives a direct relation.

The spin current result Eq. (\ref{SHjs}) indicates that the spin Hall angle $\theta_{\rm sh}$ defined by
\begin{align}
 j_{{\rm s},\beta}^\alpha =\theta_{\rm sh} \epsilon_{\alpha\beta\gamma}j_\gamma,
\end{align}
is related to $\lambda_{\rm sh}$ by $\theta_{\rm sh}=\lambda_{\rm sh}/(\sigma_{\rm e}\tau)$. 
In the present model, it is  $\theta_{\rm sh}=(\frac{4\pi}{3})^2\epsilon_{\rm so}\dos^2 \ef\simeq(\frac{4\pi}{3})^2\epsilon_{\rm so}/\ef $.

Recently, spin Hall effect was discussed in terms of spin polarization vector or a spin moment ``without spin current'' in Ref. \cite{Chen18}. 
In our context, the spin moment defined in the bulk is
\begin{align}
 P_s^{ij} \equiv \intr s_i r_j  = -\lambda_{\rm sh} \epsilon_{ijk}E_k, \label{Ps}
\end{align}
where we used our result, Eq. (\ref{SHs}), and neglected boundary contribution.
The spin moment, essentially the product of $C_{SJ}$ and $\rv$, is thus described by a uniform response function, and may be a convenient order parameter for discussing a bulk response. 
However, the concept becomes obscure in the diffusive regime, where experiments are carried out, and  simple equation like Eq. (\ref{Ps}) breaks down.
On the other hand, our local form, Eq. (\ref{SHs}), can be extended to diffusive regime as we shall see below and can describe local spin profile, which is in fact observed in experiments. 

The result (\ref{SHs}) is a constitute equation describing the response of the material when an electric field or current is applied and it should not be considered as a part of the Maxwell's equations; The equation does not mean that rotation of $\Ev$ is induced by applying a magnetic field.
Modification of the Maxwell's equations due to spin-charge conversion effects by spin-orbit interactions has been studied in Refs. \cite{Kawaguchi16,Kawaguchi18}.

\section{Diffusive regime}

We have so far discussed ballistic regime, length scale shorter than the elastic mean free path, $\ell\equiv \frac{\kf}{m}\tau$. 
Experiments are usually carried out in diffusive regime, which is theoretically considered by  including an electron ladder representing successive scattering by impurities. 
Besides normal (spin-independent) scattering, spin-orbit interaction is included in the ladder, giving rise to a decay of spin diffusion channel and a finite spin diffusion length. 
The correlation function including diffusion  is obtained by simply replacing the local correlation function by a long-ranged one (see Sec. \ref{SEC:diffusion} for details) , 
\begin{align}
 \correlation_{{\rm (D)}JS}^{\alpha\beta}(\rv,\rv')= \lambda_{\rm sh} \epsilon_{\alpha\beta i} \nabla_i 
 D_{\rm s}(\rv-\rv'),
 \label{chijsdiffresult}
\end{align}
where $D_{\rm s}(\rv)$ is diffusion propagator for electron spin, which is defined in the momentum representation by 
\begin{align}
D_{\rm s}(q) &\equiv \frac{1}{Dq^2\tau+\frac{4}{3}\gamma}. 
\end{align}
Here $\gamma\equiv \tau_0/\tau_{\rm so}$ represents the strength of spin relaxation, $\tau_{\rm so}$ being the spin lifetime due to spin-orbit interaction.

\begin{figure}
 \includegraphics[width=0.4\hsize]{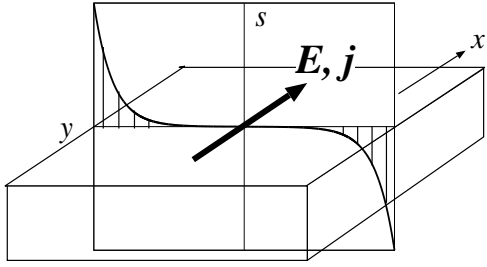}
\caption{ Schematic figure depicting spin accumulation as a result of  spin Hall effect in diffusive regime. As Eq. (\ref{SHsdiff}) indicates, the spin accumulation  formed at the edge  where $\nabla\times\jv$ is finite is smeared out by electron diffusion, resulting in a exponential profile with decay length of spin diffusion length. \label{FIGSHE}}
 \end{figure}

The spin density then reads 
\begin{align} 
\sv &= \frac{\lambda_{\rm sh}}{\sigma_{\rm e}} \intr' [\nabla D_{\rm s}(\rv-\rv')]\times  \jv(\rv') \nnr
&= \frac{\lambda_{\rm sh}}{\sigma_{\rm e}} \intr' D_{\rm s}(\rv-\rv')[\nabla\times  \jv](\rv') 
.\label{SHsdiff}
\end{align}
Obviously the resulting spin density has an exponentially decaying profile with decay length of $\ell_{\rm s}\equiv \ell/(2\sqrt{\gamma})$ at the edge where $\nabla\times\jv$ is finite, reproducing familiar spin accumulation profile of Fig. \ref{FIGSHE}.
The result (\ref{SHsdiff}) is consistent with the continuity equation (\ref{spincontinuity}) (or diffusion equation) as it is a solution of 
$\nabla\cdot\jv_{\rm s}^\alpha =s_\alpha/\tau_{\rm r}$ with $j_{{\rm s},\beta}^\alpha=D\nabla_\beta s_\alpha$ and $1/\tau_{\rm r}=4\gamma/(3\tau)$ being the spin relaxation time.

%

\section{Rashba spin-orbit interaction \label{SECRashba}}

 \begin{figure}
 \includegraphics[width=0.4\hsize]{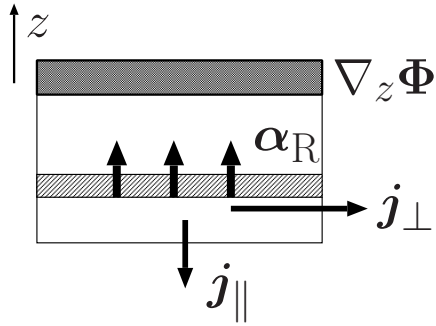}
\caption{ Schematic figure of a junction with spatially inhomogeneous spin source field $\Phiv$ and localized Rashba interaction $\alphaRv$. Conventional inverse Rashba-Edelstein effect corresponds to the current $\jv_{\perp}$ perpendicular to the junction.
\label{FIGrashbasys}}
 \end{figure}
\begin{figure}
 \includegraphics[height=3\baselineskip]{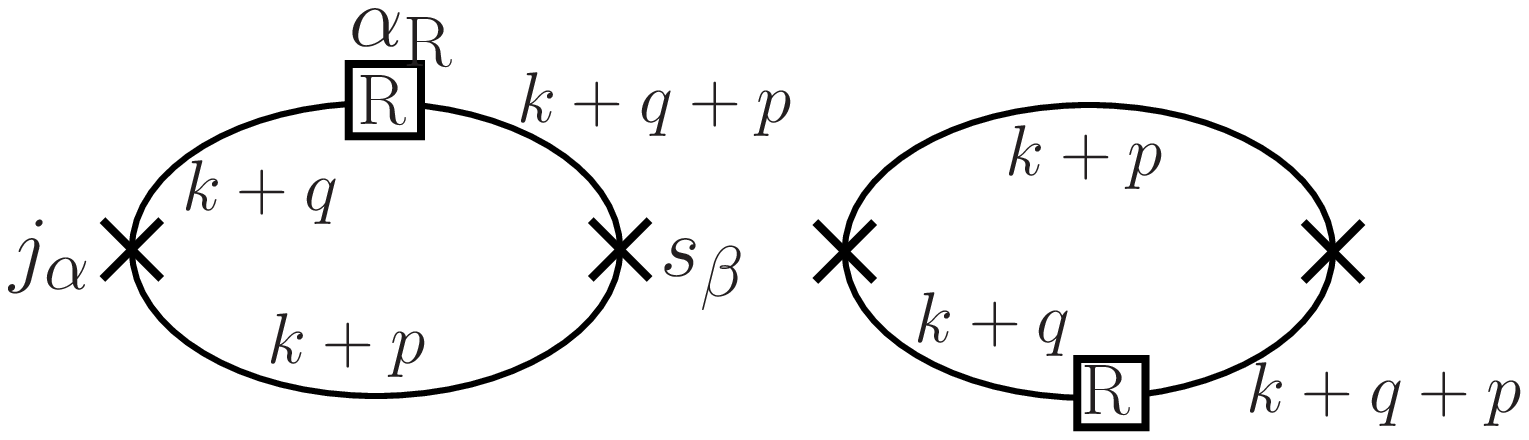}
 \includegraphics[height=3\baselineskip]{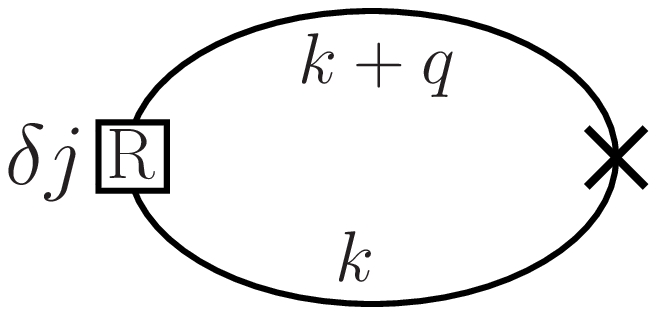}
\caption{ Diagramatic representation of current-spin correlation function including Rashba spin-orbit interaction to the linear order.  Square vertex labeled with R denotes Rashba spin-orbit interaction. The last diagram corresponds to the contribution of current correction, $\delta j$.
\label{FIGrashba}}
 \end{figure}
Let us consider the spin-charge conversion for case of the Rashba interaction, whose Hamiltonian is 
\begin{align}
 H_{\rm R} &=   - \intr c^\dagger [\alphaRv \cdot ({\pv}\times \sigmav)] c,
 \label{HsoR}
\end{align}
where $ \alphaRv$ is the Rashba vector.
Here we consider the Rashba interaction with spatial profile along the $z$ direction (Fig. \ref{FIGrashbasys}).
In experiments, Rashba interaction is localized at interfaces and thus assuming uniform 
Rashba interaction is not realistic.
The current correction is give by 
\begin{align}
\delta j_i(\rv) 
  &=  \epsilon_{ijk} \alphaR_j(\rv)  c^\dagger \sigma_k c  .
   \label{jsoi1k}
\end{align}
The current-spin correlation function for the Rashba case is diagratically depicted in Fig. \ref{FIGrashba}.
It vanishes for uniform current and uniform Rashba interaction, and the leading contribution for slowly varying case turns out to be (See Sec. \ref{SECAPPRashba})
\begin{align}
\correlation_{{\rm (R)}JS}^{\alpha\beta}
&= i 4\pi \dos D\tau^2 \lt[
 -\epsilon_{ij\beta} (\nabla_\alpha \alpha_{{\rm R},i})\nabla_j 
+\epsilon_{i\alpha \beta} (\nabla_j \alpha_{{\rm R},i})\nabla_j 
\rt].
\label{chijsssrashba}
\end{align}
Here we see that spatial modulation of the Rashba interaction generates current parallel ($j_\parallel$) and  perpendicular ($j_\perp$) to the junction, the first and the second term, respectively.
The second term corresponds to the inverse spin Hall or inverse Rashba-Edelstein effect.
The result shows that for a one-dimensional layer structure changing in the $z$ direction, the Rashba interaction acts the same as impurity spin-orbit interaction in the context of spin-charge conversion with the efficiency of 
\begin{align}
 \lambda_{\rm sh}^{\rm R}=4\pi\dos D\tau^2\nabla_z\alphaR_z. \label{lambdashR}
\end{align}
In reality, the gradient changes sign on the two different sides of the interface Rashba interaction.
For the inverse Rashba-Edelstein signal to be finite, therefore, asymmetric interface is essential. 
Obviously this fact would not be crucial, as Rashba interaction itself requires inversion symmetry breaking.

The diffusive case like with nonmagnetic spacer metal in Ref. \cite{Sanchez13} is described by including diffusion propagator in Eq. (\ref{chijsssrashba}), in the same manner as Eq. (\ref{chijsdiffresult}), as shall be mentioned in the next section.   

Conventionally, Rashba-Edelstein effect is discussed only for the uniform Rashba interaction \cite{Edelstein90}, while in experiments the interaction is localized at atomic scale at the interface. As the uniform Rashba interaction induces conversion between spin density and electric current \cite{Edelstein90,Shen14}, its gradient naturally induces gradient of spin, or 'spin current'.
The result here indicates that a spatial profile of the spin-orbit interaction leads to rich spin-charge conversion properties compared to the uniform case.

\section{Inverse spin Hall and spin pumping effects}

We here discuss spin-charge conversion effect combined with  spin pumping effect.
As shown in Ref. \cite{TataraSP17}, the spin generation field for spin pumping is  (see Sec. \ref{SECSPdetail})
\begin{align}
 \bm{\Phi}\equiv (\nv\times\dot{\nv}) \Re[\eta] + \dot{\nv}\Im[\eta],
\end{align}
where $\nv$ is a unit vector denoting magnetization direction and 
a complex parameter $\eta$ is written in terms of spin-dependent hopping amplitude of electron across the interface, $\tilde{t}_\sigma$ ($\sigma=\pm$ denotes the spin) as
$\eta=\frac{\chi_0}{2} \tilde{t}_+^* \tilde{t}_- $ with $\chi_0$ is uniform susceptibility. 
The field induces an imaginary part of the lesser component of the N electron self energy, 
\begin{align}
 \Sigma_{\rm N}^< = i\sigmav \cdot \bm{\Phi} \label{SigmaN}.
\end{align}
For metallic junctions, the spin source field $\bm{\Phi}$ is localized at the interface and decays rapidly away from the interface. 
The generated electric current is 
\begin{align}
 j_{\alpha}(\rv) = \intr' \correlation_{JS}^{\alpha\beta}(\rv-\rv') \Phi_\beta(\rv') .
 \label{jish0}
\end{align}
For ballistic electron transport using Eq. (\ref{chijsresult}), the inverse spin Hall current is given by a derivative of the spin source field $\bm{\Phi}$ as 
\begin{align}
 j_{\alpha}= \lambda_{\rm sh} \epsilon_{\alpha\beta i} \nabla_i \Phi_\beta .
 \label{jish1}
\end{align}
Note that all the coefficients in the current are microscopically defined in the present model.
The geometry of the current and magnetization agrees with conventional inverse spin Hall phenomenology \cite{Saitoh06}. 
According to our result, the total current integrated over the thickness of N layer is directly related to the magnitude of spin accumulation (or spin source field) at the interface. Choosing the interface as in the  $xy$ plane, the total current in the ballistic regime is 
\begin{align}
\Iv\equiv \int dz \jv = \lambda_{\rm sh} \biggl[ \zvhat\times 
\biggl( \Re[\eta](\nv\times\dot{\nv})  + \Im[\eta]\dot{\nv}\biggr) \biggr], \label{Itotalballistic}
\end{align}
where $\zvhat$ is a unit vector in the $z$ direction.

In the diffusive regime, the correlation function $\correlation_{JS}^{\alpha\beta}$ becomes a diffusive one, obtained by replacing $q_i$ in Eq. (\ref{chijsresult}) by $q_i D_{\rm s}(q)$ (see Eq. (\ref{chijsdiffresult})).
The inverse spin Hall current due to spin pumping effect then is 
\begin{align}
 j_{\alpha}^{\rm (D)} (\rv) = \lambda_{\rm sh} \epsilon_{\alpha\beta i} \nabla_i \intr'  D_{\rm s}(\rv-\rv') \Phi_\beta(\rv') .
 \label{jishD}
\end{align}
Using $D_{\rm s}(\rv)=a\int\frac{dq}{2\pi}\frac{e^{iqr}}{Dq^2\tau+\kappa}=\frac{3a \ell_{\rm s}}{2\ell^2}e^{-r/\ell_{\rm s}}$ 
($a$ is the lattice constant, $\kappa\equiv \frac{4}{3}\gamma$ and $\ell_{\rm s}=\ell/\sqrt{3\kappa}$ is spin diffusion length), 
the total current is $\frac{3\ell_{\rm s}a}{2\ell^2}$ times $I$ of ballistic value (Eq. (\ref{Itotalballistic})).
The case of Rashba interaction is described by replacing the coefficient $\lambda_{\rm sh}$ by $\lambda_{\rm sh}^{\rm R}$ of Eq. (\ref{lambdashR}).

The long-ranged current of Eq. (\ref{Itotalballistic}) indicates that there is a spin motive force which propagates through normal metal. 
In the case of spatially uniform Rashba interaction and magnetization, the local motive force was found to be 
$\Ev_{\rm R} \propto [\alphaRv\times\dot{\nv}+\beta \alphaRv\times(\nv\times\dot{\nv})]$, where $\beta$ is a constant representing spin relaxation \cite{Kim12,Tatara_smf13,Nakabayashi14}.
The expression for the current Eq. (\ref{jishD}) indicates that there is a long-ranged version of Rashba-induced spin motive force mediated by electron diffusion. 
The counterpart of the long-ranged spin motive force, namely, the long-ranged spin Berry's phase, has been pointed out in the context of anomalous Hall effect \cite{Nakazawa14}.

\section{Combination of inverse and direct spin Hall effects}

%
\begin{figure}
\includegraphics[width=0.4\hsize]{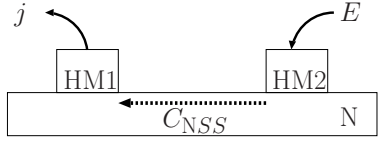}
\caption{Schematic figure of nonlocal direct and inverse spin Hall experiment. The electric field is applied to a heavy metal HM2, resulting in 'spin current generation' by direct spin Hall effect. The spin signal is transmitted through nonmagnetic metal (N) to another heavy metal (HM1) and induces electric current due to the inverse spin Hall effect. Conventional picture of the coupling between HM1 and HM2 is due to spin current propagation, while in the present picture, they are connected by spin correlation function $C_{{\rm N}SS}$ of N. 
\label{FIGDISH}}
 \end{figure}
Here we discuss the inverse spin Hall effect in a heavy metal (HM1) induced nonlocally by spin Hall effect in another heavy metal (HM2) separated by thin nonmagnetic metal (N) (Fig. \ref{FIGDISH}) (nonlocal spin detection setup).  
The observed current is written by the full lesser Green's function, ${\cal G}^{<}_{\rm HM1}$, containing all the interactions of HM1 as 
\begin{align}
 j_i(\rv)&=-i\tr[v_i {\cal G}^{<}_{\rm HM1}(\rv,\rv)],
\end{align}
where $\rv$ is a position in HM1.
We include the effect of the electric field applied to HM2 using a vector potential $\Av$.
Denoting the (path-ordered) Green's function connecting HM1 and position $\rv'$ at HM2 as $G(\rv,\rv')$, the current at the linear response reads 
\begin{align}
 j_i(\rv)&=-i \intr' \tr[v_i {G}(\rv,\rv') v_j G(\rv',\rv)]^< A_j(\rv'),
\end{align}
where the integral of $\rv'$ is over HM2.
The connection between HM1 and HM2 in the Green's function $ G(\rv',\rv)$ arises from N electrons. 
Writing the electron hopping amplitude across the interfaces by a constant $t$,  $ G(\rv',\rv)$ is decomposed  in the perturbative regime into a product of Green's functions for HM1, N and HM2 (denoted by $G_{\rm HM1}$, $G_{\rm N}$ and $G_{\rm HM2}$, respectively) as
\begin{align}
 G(\rv,\rv') &=t^2 \int_{I1} d\rv_1  \int_{I2} d\rv_2 G_{\rm HM1}(\rv,\rv_1)G_{\rm N}(\rv_1,\rv_2)G_{\rm HM2}(\rv_2,\rv'),
\end{align}
where $\rv_i$ ($i=1,2$) is position at interface (I$i$) between NM$i$ and N. 

\begin{figure}
\includegraphics[height=3\baselineskip]{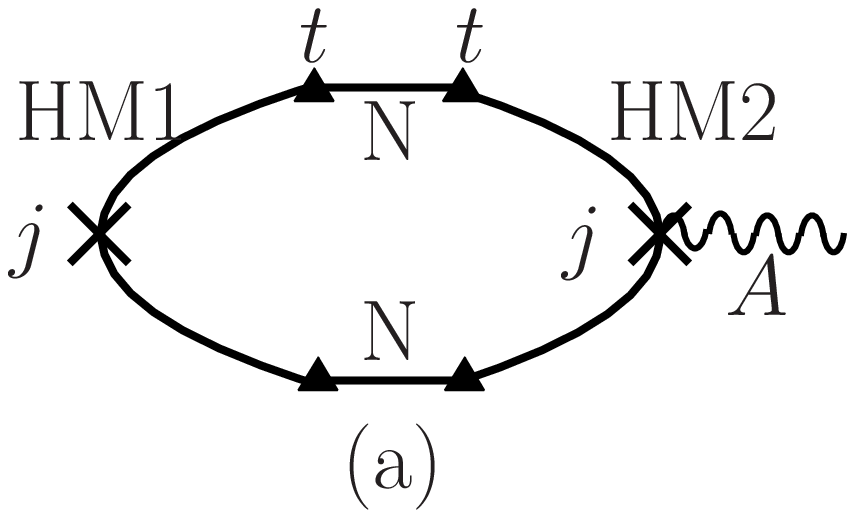}
\includegraphics[height=3\baselineskip]{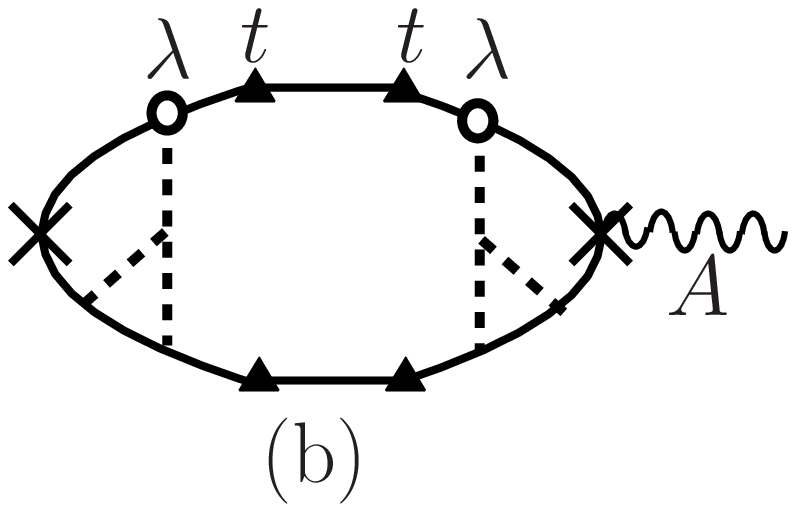}
\includegraphics[height=3\baselineskip]{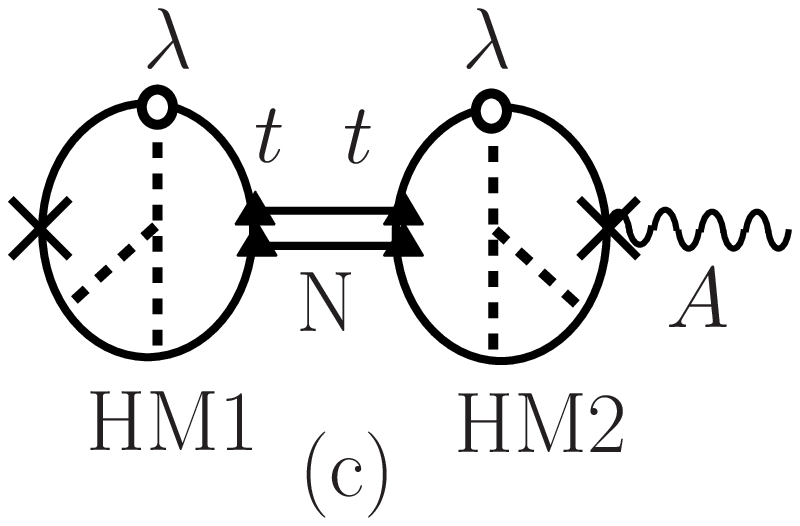}
\caption{ (a): Nonlocal current correlation function $C_{JJ}$ connecting HM1 and HM2. Triangular vertex denotes the hopping between HM and N with amplitude $t$. 
(b): $C_{JJ}$ with impurity spin-orbit interaction explicitly shown to the lowest order. Complex congugate processes like in Fig. \ref{FIGSCC} are not shown.
(c) is a diagram where propagator in N carrying small wave vector is suppressed. The contributions of HM1 and HM2 reduce to  the correlation discussed for spin Hall effect. 
\label{FIGCjjnl}}
 \end{figure}

Let us now calculate the lesser component following the standard treatment of vector potential, namely, it carries infinitesimal angular frequency $\Omega$ and consider the limit of $\Omega\ra0$. Other interactions are treated as static and do not change the electron angular frequency.
From the above argument, we have (suppressing spatial coordinates) 
\begin{align}
 {\cal G}^{<}_{\rm HM1} &= t^4 \sumom [ G_{\rm HM1}(\omega)G_{\rm N}(\omega)G_{\rm HM2}(\omega) v_j 
 G_{\rm HM2}(\omega+\Omega)G_{\rm N}(\omega+\Omega)G_{\rm HM1}(\omega+\Omega) ]^< \nnr
 &\simeq  t^4 \sumom (f(\omega+\Omega)-f(\omega))  [G_{\rm HM1}(\omega)G_{\rm N}(\omega)G_{\rm HM2}(\omega)]^\ret v_j \nnr
& \times 
[ G_{\rm HM2}(\omega+\Omega)G_{\rm N}(\omega+\Omega)G_{\rm HM1}(\omega+\Omega)]^\adv A_j(\Omega) ,
\end{align}
which at low temperatures reduces to (using $f'(\omega)\simeq -\delta(\omega)$ and $-i\Omega A_j=E_j$) 
\begin{align}
 {\cal G}^{<}_{\rm HM1} &= i  t^4  [G_{\rm HM1}G_{\rm N}G_{\rm HM2}]_{\omega=0}^\ret (\vv\cdot\Ev)  
[ G_{\rm HM2}G_{\rm N}G_{\rm HM1}]_{\omega=0}^\adv ,
\end{align}
where all the Green's functions are at zero angular frequency ($\omega=0$).
We thus have linear response formula for the generated current as 
\begin{align}
 j_i(\rv)=\intr' C_{JJ}^{ij}(\rv,\rv') E_j(\rv'),
\end{align}
where 
\begin{align}
C_{JJ}^{ij}(\rv,\rv') &\equiv i  t^4  \tr[ v_i [G_{\rm HM1}G_{\rm N}G_{\rm HM2}]_{\omega=0}^\ret v_j   
[ G_{\rm HM2}G_{\rm N}G_{\rm HM1}]_{\omega=0}^\adv ],
\end{align}
is a nonlocal current correlation function connecting $\rv\in$HM1 and $\rv'\in$HM2. 
The correlation function is schematically shown in Fig. \ref{FIGCjjnl}(a).

The Green's functions $G_{{\rm HM}i}$ contain spin-orbit interaction, which we now treat perturbatively for the case of impurity spin-orbit interaction.
We thus include for both HM1 and HM2 the 'spin-charge conversion' vertex, combination of spin-orbit interaction and normal impurity scatterings, as in Fig. \ref{FIGSCC} (Fig. \ref{FIGCjjnl}(b)). 
We consider long-range (diffusive) transmission through N to describe experimental sistuations.
Then the N propagator carries only a small wave vector, and the HM verteces are evaluated for small incoming wave vector $q$, resulting in a diagram of Fig. \ref{FIGCjjnl}(c), where spin-charge conversion in HMs are detemined locally.
We thus have 
\begin{align}
C_{JJ}^{ij} &= i (\lambda_{\rm sh})^2 t^4 \epsilon_{i\alpha\beta}\epsilon_{j\alpha'\beta'}
\nabla_{\beta}^{(\rv)} \nabla_{\beta'}^{(\rv')} C_{{\rm N}SS}^{\alpha\alpha'}(\rv,\rv'), \label{Cjj1}
\end{align}
where $\nabla_{\beta}^{(\rv)}$ denotes a derivative with respect to $\rv$ and 
\begin{align}
 C_{{\rm N}SS}^{\alpha\alpha'} (\rv,\rv')
 &\equiv \tr[  \sigma_\alpha G_{\rm N}^\ret(\rv,\rv') \sigma_{\alpha'} G_{\rm N}^\adv(\rv',\rv)],
\end{align}
is  the spin correlation function of N electron connecting HM1 and HM2.
For nomnagnetic metal we consider, $C_{{\rm N}SS}^{\alpha\alpha'}$ is spin diagonal.
Including electron diffusion in N, we have 
\begin{align}
 C_{{\rm N}SS}^{\alpha\alpha'}(\rv,\rv') &= \delta_{\alpha\alpha'}
 2\pi\dos\tau D_{\rm s}(\rv-\rv').
\end{align}
The generated current is therefore represented by the correlation function
\begin{align}
C_{JJ}^{ij}(\rv,\rv') &=- i 2\pi\dos\tau (\lambda_{\rm sh})^2 t^4 
(\delta_{ij}\nabla^2 -\nabla_{i} \nabla_{j}) D_{\rm s}(\rv-\rv'), \label{Cjjres}
\end{align}
where $i$ and $j$ are directions of generated current and applied electric field, respectively, and $\nabla$ acts on $\rv$. 
Equation (\ref{Cjjres}) applies only for the in-plane current directions $i,j$, because derivative in the perpendicular direction of N propagator vanishes in Eq. (\ref{Cjj1}).
The result (\ref{Cjjres}) reproduces conventional nonlocal 'spin current' detection scheme without ambiguity.

The present approach can straightforwardly extended to the case of other spaces such as antiferromagnet and insulators, simply replacing the N propagator $C_{{\rm N}SS}$ by the corresponding spin correlation function of the material.
While we have weak temperature dependence in the present metallic case because of large Fermi energy, rich dependence is expected  when magnons or phonons dominate spin transport.

\section{Summary and discussion}
We have presented a unified formalism to describe spin-charge conversion phenomena in terms of correlation functions between physical observables, spin and electric current, without referring to spin current.
Considering simple theoretical models for the case of metals, we derived concise formula for spin Hall effect and the inverse spin Hall effect combined with spin pumping and spin Hall effects, avoiding ambiguities associated with spin current. 
While this approach describes already well-known physical processes, this here developed pictures has been disregarded so far, possibly due to the at first sight very accessible description using spin currents (which however entail ambiguities).

From the viewpoint of linear response theory, the difference between the responses of spin current  and spin density when an electric field is applied is that the former is represented by the  the uniform ($q=0$) component ($q$ is the external wave vector) of a response function (correlation function), while the latter requires $q$-resolved response ($q$-linear component).
Instead, the latter directly provides physical quantity, while the spin current needs to be mapped to spin density at the end.
In this context, spintronics effects provide us novel useful tools to access $q$-resolved magnetic susceptibility by use of electric measurements.

In our representation, the spin current transmission turned out to be the propagation or correlation of spin fluctuation, and thus its efficiency is determined essentially by the (spatial derivative of) magnetic susceptibility.
This is not surprising in the case of charge transport: The electric conductivity, $\sigma_{\rm e}$, represented by correlation of electric current, is expressed in terms of electric permittivity, $\varepsilon$, which is a correlation of electric density, as $\sigma_{\rm e}=i\varepsilon_0\omega(1-\varepsilon)$ at finite angular frequency $\omega$, indicating that electric current is mediated by charge fluctuation.
This relation is as a result of charge conservation or gauge invariance.
In the case of spin transport, however, such a universal relation does not exist, and thus studying spin current propagation does not itself give physical response.
What we proposed here is to describe spin density directly within the linear response theory, instead of using spin current and applying classical diffusion equation to evaluate physical spin density.

Recently, 'spin transmission' through antiferromagnetic insulators \cite{Qiu16} and metals \cite{Qu15} have been studied  experimentally and theoretically \cite{Cramer18}, and intriguing features have been reported. Extention of the present scheme to cover those systems are under way.

\acknowledgements
GT thanks E. Saitoh, T. Ono, A. Shitade, J. Fujimoto, H. Kohno, J. Shibata, A. Takeuchi
and M. Kl\"aui
for valuable discussions.
This work was supported by 
a Grant-in-Aid for Exploratory Research (No.16K13853) and a Grant-in-Aid for Scientific Research (B) (No. 17H02929) from the Japan Society for the Promotion of Science,
a Grant-in-Aid for Scientific Research on Innovative Areas (No.26103006) from The Ministry of Education, Culture, Sports, Science and Technology (MEXT), Japan, 
and 
the Graduate School Materials Science in Mainz (MAINZ) (DFG GSC 266).

\appendix
\section{Correlation function $C_{SJ}$ and physical response \label{SEC:correlation}}
Here  relation between correlation function $C_{SJ}$ and physical quantity induced at the linear response to an external field is summarized.
We demosntrate the case of spin Hall effect, described in terms of spin accumulation, which reads
\begin{align}
 s_\alpha(\rv,t) &= -i\tr[\sigma_\alpha G^<(\rv,t,\rv,t)],
\end{align}
where $G^<$ is the lesser component of the full Green's function including all the interactions and $\alpha=x,y,z$ denotes spin durection. 
The driving field for spin Hall effect is an external electric field, $\Ev=-\dot{\Av}$, where $\Av$ is a vecor potential.
The interaction Hamiltonian of the vector potential reads 
$H_{A}=\intr \Av\cdot \jv$, where $\jv\equiv c^\dagger {\vv} c$, ${\vv}$ is the velocity vector operator. 
Including the vector potential perturbatively to the linear order in $G^<$, we obtain 
the Fourier representation ($ s_\alpha(\rv,t) =\sumOm\sum_{\qv}e^{i(\qv\cdot\rv-\Omega t)} s_\alpha(\qv,\omega)$) as 
\begin{align}
 s_\alpha(\qv,\omega) &= -i \chi_{SJ}^{\alpha\beta}(\qv,\Omega) A_\beta(\qv,\Omega),
\end{align}
where 
\begin{align}
 \chi_{SJ}^{\alpha\beta}(\qv,\Omega) &= \sumom\sum_{\kv} \tr[\sigma_\alpha g_{\kv+\qv,\omega+\Omega} {v}_\beta g_{\kv\omega}]^<,
\end{align}
where $g_{\kv\omega}$ is the path-ordered Green's function without the external field having wave vector $\kv$ and angular frequency $\omega$. 
The lesser component is decomposed into the retarded and advanced Green's functions, $g^\ret$ and $g^\adv$, respectively, resulting in
\begin{align}
 \chi_{SJ}^{\alpha\beta}(\qv,\Omega) = & \sumom\sum_{\kv}  \tr\biggl[ \sigma_\alpha \biggl[
-(f(\omega+\Omega)-f(\omega))  g^\ret_{\kv+\qv,\omega+\Omega} {v}_\beta g^\adv_{\kv\omega} \nnr
& +f(\omega+\Omega)   g^\adv_{\kv+\qv,\omega+\Omega} {v}_\beta g^\adv_{\kv\omega} 
-f(\omega) g^\ret_{\kv+\qv,\omega+\Omega} {v}_\beta g^\ret_{\kv\omega} \biggr]
\biggr],
\end{align}
where $f(\omega)\equiv [e^{\beta\omega}+1]^{-1}$ is the Fermi distribution function. 
The term containing $f(\omega+\Omega)-f(\omega)$ is the  Fermi surface contribution representing the dominant electron excitation, while other two terms called the Fermi sea contribution represent mostly equilibrium effects with small correction to the Fermi surface contribution. 
We consider the simple case where Fermi sea contribution is negligible.
For slowly-varying external field, we expand with respect to $\Omega$ and retain the linear order.
The result is, using $f'(\omega)=-\frac{\beta}{4} [\cosh\frac{\beta\omega}{2}]^{-2}\simeq -\delta(\omega)$ valid at low temperatures,  
\begin{align}
 \chi_{SJ}^{\alpha\beta}(\qv,\Omega) &= -i\Omega C_{SJ}^{\alpha\beta}(\qv) ,
\end{align}
where 
\begin{align}
 C_{SJ}^{\alpha\beta}(\qv) \equiv \sum_{\kv}  \tr [ \sigma_\alpha  g^\ret_{\kv+\qv} {v}_\beta g^\adv_{\kv}], 
\end{align}
is the correlation function defined in Eq. (\ref{correlationdef}), with $g^\adv_{\kv}\equiv g^\adv_{\kv,\omega=0}$.
We thus have
\begin{align} 
s_\alpha
&=  C_{SJ}^{\alpha\beta}(\qv) E_\beta,
\end{align}
indicating that the  correlation function $C_{SJ}^{\alpha\beta} $ represents the dominant Fermi surface contribution of the spin density induced by external electric field (spin Hall effect).

\section{Calculation of diffusion propagators \label{SEC:diffusion}}
 \begin{figure}
 \includegraphics[width=0.4\hsize]{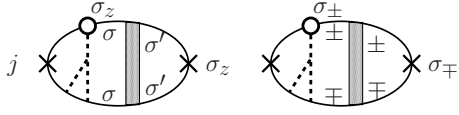}
\caption{ Feynmann diagram describing the inverse spin Hall effect at the linear order in the impurity spin-orbit interaction in the diffusie regime. Cases of spin vertices $\sigma_z$ and $\sigma_\mp$ are separately shown, with $\sigma$, $\pm$ representing spin indices. 
\label{FIGISHdiff}}
 \end{figure}
The diffusive correction to the inverse spin Hall effect is diagrammatically represented in Fig. \ref{FIGISHdiff}, where the hatched square denotes  a ladder  of  successive scatterings by impurities and impurity spin-orbit interaction. 

\begin{figure}
 \includegraphics[width=0.5\hsize] {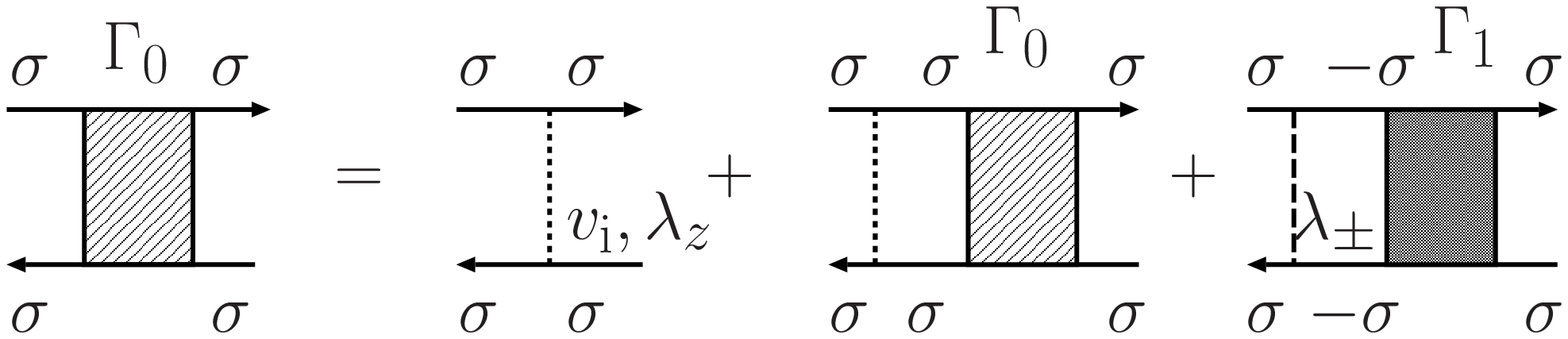}
 \includegraphics[width=0.5\hsize] {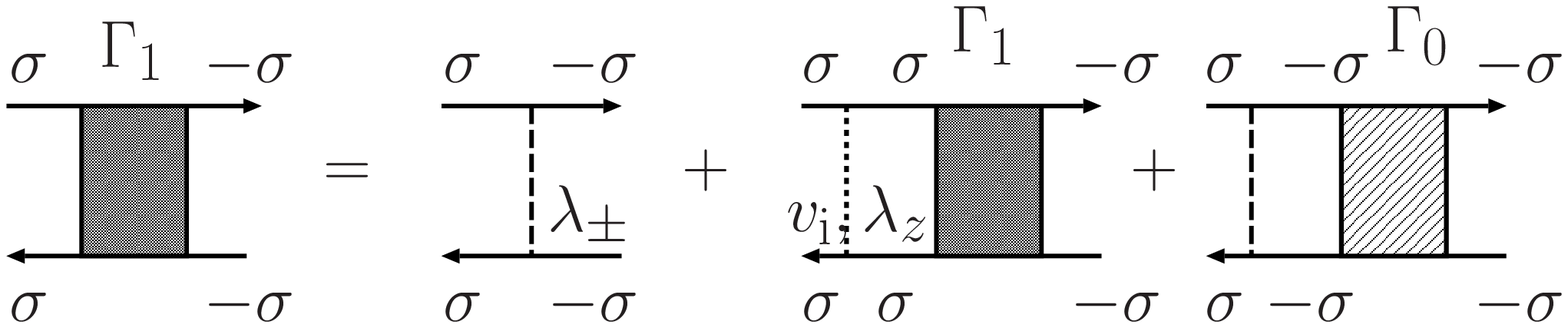}
 \includegraphics[width=0.5\hsize] {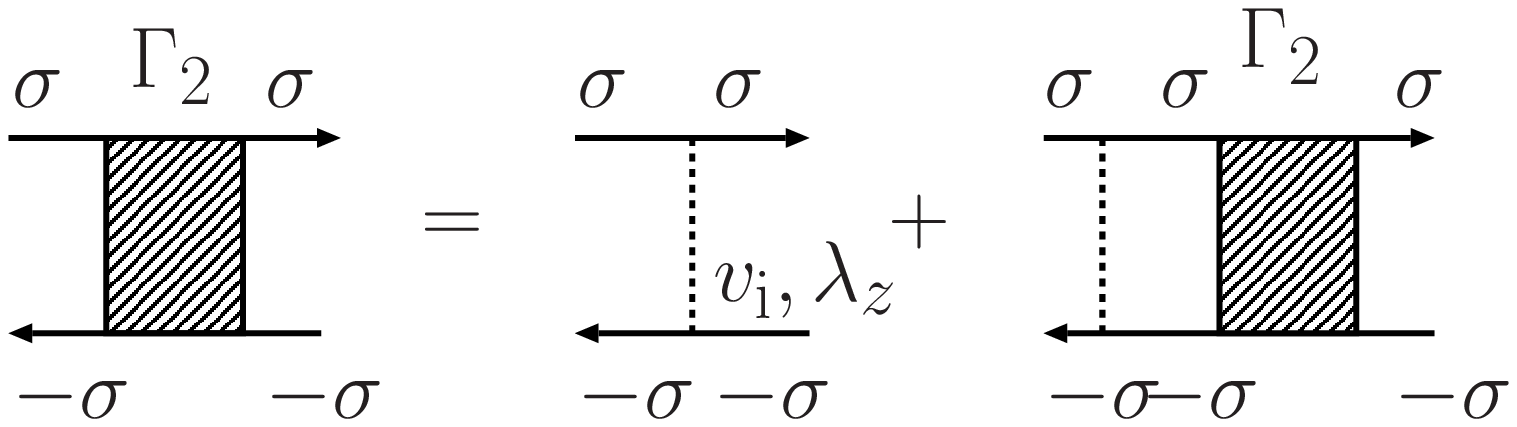}
\caption{ Diagramatic representation of impurity ladders representing diffusive motion.
\label{FIGdiffusionsoi}}
\end{figure}

The diffusion ladder is calculated as follows.
The electron elastic lifetime $\tau$ is thus given by $1/\tau=1/\tau_0+1/\tau_{\rm so}$, where
$\tau_0$ is the lifetime due to nonmagnetic impurities and $1/\tau_{\rm so}=\frac{4\pi}{3}\lambda^2 \Vi^2 \dos \kf^4$ is  due to spin-orbit impurity.
The elastic scattering rate $\frac{1}{\tau_{\rm so}}$ includes spin-conserving and spin flip contributions, $\frac{1}{\tau_{\rm so}^z}$ and $\frac{1}{\tau_{\rm sf}}$, respectively. For the present isotropic model, we have 
$\frac{1}{\tau_{\rm sf}}=\frac{2}{3} \frac{1}{\tau_{\rm so}}$.
Assuming that each pair of two scatterings (represented by a line connecting retarded and  advanced Green's functions) conserves spin, we have three diffusion ladders with different spin indices to take account of, $\Gamma_i$ ($i=0,1,2$). 
$\Gamma_0$ and $\Gamma_1$ correspond to the ladder for diagonal spin vertex ($\sigma_z$) shown in Fig. \ref{FIGdiffusionsoi}(a) with spin indices $\sigma'=\sigma$ and $\sigma'=-\sigma$, respectively.  
As is seen easily, the two cases  $\sigma'=\sigma$ and $\sigma'=-\sigma$ have different signs, and thus the diffusion for diagram (a) is $\Gamma_0-\Gamma_1$.  
For off-diagonal spin vertices, $\sigma_\pm$, contributing diagram is  Fig. \ref{FIGdiffusionsoi}(b), with diffusion ladder $\Gamma_2$.
These ladders satisfy coupled equations derived perturbatively   (see Fig. \ref{FIGdiffusionsoi} for diagramatic representation)
\begin{align}
 \Gamma_0 &= V_{\rm i} \biggl[
   \lt(1+\frac{\gamma}{3}\rt)\lt(1+\Pi\Gamma_0\rt) +\frac{2\gamma}{3} \Pi \Gamma_1 \biggr] \nnr
 \Gamma_1 &= V_{\rm i} \biggl[
   \frac{2\gamma}{3}\lt(1+\Pi\Gamma_0\rt) +\lt(1+\frac{\gamma}{3}\rt) \Pi \Gamma_1 \biggr] \nnr
 \Gamma_2 &= V_{\rm i} 
   \lt(1-\frac{\gamma}{3}\rt)\lt(1+\Pi\Gamma_2\rt),
\end{align}
where $\gamma\equiv \frac{\tau_0}{\tau_{\rm so}}$, $V_{\rm i}\equiv (2\pi\dos\tau_0)^{-1}$ represents square of normal impurity potential.
Function $\Pi(q)$, representing an particle-hole pair propagation, is evaluated for small $q/\kf$ as 
$\Pi(q)\equiv \sumkv g^\ret_{\kv+\frac{\qv}{2}} g^\adv_{\kv-\frac{\qv}{2}}=2\pi\dos\tau(1-Dq^2\tau)=\frac{1}{V_{\rm i}}(1-\gamma)(1-Dq^2\tau)$, where $D\equiv \frac{\kf^2}{3m^2}\tau$ is the diffusion constant.
The solutions are 
\begin{align}
 \Gamma_0 &= V_{\rm i} \frac{1}{2}[D(q)+D_{\rm s}(q)] \nnr
 \Gamma_1 &= V_{\rm i} \frac{1}{2}[D(q)-D_{\rm s}(q)] \nnr
 \Gamma_2 &= V_{\rm i} D_{\rm s}(q),
 \end{align}
where 
\begin{align}
D(q)&\equiv \frac{1}{Dq^2\tau} \nnr
D_{\rm s}(q) &\equiv \frac{1}{Dq^2\tau+\frac{4}{3}\gamma}, 
\end{align}
represent charge and spin diffusion, respectively.
As a result, all the diffusions contributing to the inverse spin Hall effect turn out to be the spin diffusion, $D_{\rm s}(q)$, which decays with decay length of  spin diffusion length, $\ell_{\rm sf}\equiv \ell/(2\sqrt{\gamma})$.
Of experimental interest is the uniform component of the current averaged over the $xy$ plane. The diffusion propagator then reduces to a one-dimensional one, 
\begin{align}
 D_{\rm s}=\int \frac{dq}{2\pi}\frac{e^{iqz}}{Dq^2\tau+\gamma}=\frac{3}{2}\frac{\ell_{\rm s}}{\ell^2}e^{-z/\ell_{\rm sf}}.
\end{align}

%
%
\section{Summary of spin pumping effect \label{SECSPdetail}}
Here we summarize the theoretical description of spin pumping effect of metallic ferromagnet (FM) in contact with a nomral metal (N). 
Conventionally, spin pumping effect has been discussed in terms of generated spin current in the normal metal, which is not a physical observable. 
Here we focus only on physical observables following the field-theoretical discussion of Ref. \cite{TataraSP17}.
What is necessary is then simply the lesser Green's function of N electron, $G_{\rm N}^<$, including all the interactions and the effect of ferromagnet.
The effect of the ferromagnetic metal arises from the interface hopping of the conduction electron between FM and N, represented by Hamiltonian of 
\begin{align}
 H_{\rm I} & \equiv \sum_{ij\in{\rm I}} 
 \lt( 
c^\dagger_i  t  d_j + 
  d^\dagger_j t^* c_i \rt),
 \label{FNhoppingreal}
\end{align}
where $i$ and $j$ denote atomic site at the interface (I) of N and FM, respectively  and $c$ and $d$  are field operators of N and FM electron, respectively.
The hopping amplitude $t$ is generally a complex $2\times2$ matrix in spin space.
Moreover, in the laboratory frame, it is time-dependent and has off-diagonal components because of dynamics magnetization.  
Including the effect of ferromagnet as a self energy $\Sigma_{\rm N}$, the lesser Green's function can be formally solved in terms of retarded and advanced components  as
\begin{align}
 G_{\rm N}^<=(1+G^\ret_{\rm N}\Sigma_{\rm N}^\ret) g^<_{\rm N}(1+\Sigma_{\rm N}^\adv G^\adv_{\rm N}) 
 + G^\ret_{\rm N}\Sigma_{\rm N}^<G^\adv_{\rm N},
 \label{DysonGNsol}
\end{align}
while the first term on the right-hand side is irrelavant, resulting in 
\begin{align}
 G_{\rm N}^<
= G^\ret_{\rm N}\Sigma_{\rm N}^<G^\adv_{\rm N},
 \label{DysonGNsol1}
\end{align}

The self energy is calculated as follows.
Conduction electron in the ferromagnetic metal is represented by the Hamiltonian
\begin{align}
 H_{\rm F} &\equiv \int_{\rm F}d^3r d^\dagger
   \lt( -\frac{\nabla^2}{2m}-\ef -\spol \nv(t)\cdot \sigmav \rt) d ,
\end{align}
where $\nv(t)$ is a unit vector representing time-dependent magnetization  direction, and the integral $\int_{\rm F}d^3r$ is within the ferromagnet.
To treat the dynamic magnetization, we switch to a rotating frame by carrying out a unitary transformation to diagonalize the $sd$ exchange interaction.
FM electron operator in the rotating frame is defined as 
\begin{align}
 \tilde{d}(t)\equiv U(t)^{-1} d(t),
\end{align}
where $U(t)$ is a time-dependent $2\times2$ unitary matrix for spin. 
The Hamiltonian in the rotating frame reads 
\begin{align}
 H_{\rm F} &= \sumkv \dtil_\kv^\dagger 
  \lt( \begin{array}{cc} 
         \epsilon_{k}-\spol -\Ascal{t}{z}  & \Ascal{t}{-} \\
                    \Ascal{t}{+}  & \epsilon_{k}+\spol +\Ascal{t}{z} 
        \end{array} \rt) \dtil_\kv ,
        \label{HF2}
\end{align} 
where $\Ascal{t}{\alpha}\equiv \frac{-i}{2}\tr[ \sigma_\alpha U^{-1}\partial_t U]$ ($\alpha=x,y,z$ denotes the spin index) is the time-component of an SU(2) gauge field, or spin chemical potential induced dynamically. 
The $z$-component of the gauge field is the adiabatic component, and is neglected to the lowest order in the spin pumping effect.
As the $sd$ coupling is diagonalized by the transformation, the hopping amplitude in the rotating representation is also diagonal, namely,
\begin{align}
\tilde{t} 
& \equiv U^{-1} t U =
\lt(\begin{array}{cc} 
\tilde{t}_+  & 0 \\ 0 & \tilde{t}_-
\end{array}\rt),
\end{align}
but the hopping intearction includes a matrix $U^{-1}$ as $c^\dagger_i  t  d_j=c^\dagger U^{-1}\tilde{t}. \tilde{d}$
Amplitude $\tilde{t}_\pm$ is the one measured in experiments for static and uniform magnetization. 
The self energy at the linear order in the  nonadiabatic gauge field reads 
\begin{align}
{\Sigma}_{\rm N}^<&= \sumom \sumkv U^{-1} \tilde{t} [g_{{\rm F}\kv}(\omega)(\Ascal{t}{} \cdot \sigmav) g_{{\rm F}\kv}(\omega)]^< \tilde{t}^* U,
\end{align}
where $g_{{\rm F}\kv}$ is the spin-polarized ($2\times2$ matrix) Green's function of FM. 
Here we have an integral over angular frequency $\omega$ as a result of an approximation. 
In fact, the integral originally acts on all the Green's functions including N electron. 
In reality, however, the integral does not have essential effects as the low energy transport is dominated by the zero frequency contribution, i.e., near the Fermi surface of N electron. We thus evaluate the integral only for Fm electron setting $\omega=0$ for N electron Green's functions.   
The self energy is calculated noting that $\omega$-integral vanishes for components containing only either retarded or advanced Green's functions as  
\begin{align}
\sumom \sumkv [g_{{\rm F}\kv}(\omega)(\Ascal{t}{} \cdot \sigmav) g_{{\rm F}\kv}(\omega)]^<
&= 
\sumom \sumkv \sum_{\pm}\Ascal{t}{\mp} (f_{\kv\mp}-f_{\kv\pm}) g^\ret _{{\rm F}\kv\pm}(\omega) g_{{\rm F}\kv\mp}^\adv(\omega) \sigma_{\pm} \nnr
&=-i\chi_0 \sum_{\pm}\Ascal{t}{\mp} \sigma_\pm,
\end{align}
where $f_{\kv\pm}$ is the Fermi distribution function of FM conduction electron with spin $\pm$ and
\begin{align}
\chi_0&\equiv - \sumkv\frac{f_{\kv\mp}-f_{\kv\pm}}{\epsilon_{\kv\mp}-\epsilon_{\kv\pm}+i\eta} 
=\frac{n_+-n_-}{2\spol}, 
\end{align}
is uniform static susceptibility ($n_\pm$ is spin-resolved electron density).
The diagonal hopping amplitude with Pauli matrix reduces to 
$ \tilde{t} \sigma_\pm \tilde{t}^*= \sigma_\pm (\Re[T_{\uparrow\downarrow}]\mp i \Im[T_{\uparrow\downarrow}])$, where 
$T_{\uparrow\downarrow}\equiv \tilde{t}_+^* \tilde{t}_-$.
The self energy then is
\begin{align}
{\Sigma}_{\rm N}^<&= -i\chi_0  U^{-1} [ \Re[T_{\uparrow\downarrow}] \Ascal{t}{\perp} +\Im[T_{\uparrow\downarrow}] (\zvhat\times \Ascal{t}{\perp})]\cdot \sigmav U,
\end{align}
where $\Ascal{t}{\perp}\equiv (\Ascal{t}{x},\Ascal{t}{y},0)$.
Using $  U^{-1}\sigma_i U={\cal R}_{ij}\sigma_j$, where ${\cal R}$ is the rotation matrix element \cite{TataraSP17}, and 
\begin{align}
 {\cal R}_{ij}(\Ascalv{t}^{\perp})_j &= -\frac{1}{2}(\nv\times \dot{\nv})_i \nnr
 {\cal R}_{ij}(\zvhat\times \Ascalv{t}^{\perp})_j &= -\frac{1}{2} \dot{\nv}_i ,
\end{align}
we finally obtain 
\begin{align}
{\Sigma}_{\rm N}^<&= i\frac{\chi_0}{2}  [ \Re[T_{\uparrow\downarrow}](\nv\times\dot{\nv}) +\Im[T_{\uparrow\downarrow}] \dot{\nv}]\cdot \sigmav \equiv i\bm{\Phi}\cdot\sigmav.
\end{align}
The spin pumping effect is therefore represented by a dynamically-induced interface spin source for N electron, 
\begin{align}
\Phiv = \frac{\chi_0}{2}  [ \Re[T_{\uparrow\downarrow}](\nv\times\dot{\nv}) +\Im[T_{\uparrow\downarrow}] \dot{\nv}].
\end{align}

The case of insulator ferromagnet was studied in Ref. \cite{Ohnuma14}.

\section{Calculation of Rashba effect \label{SECAPPRashba}}
%
In this section, details of the calculation in Sec. \ref{SECRashba} are shown.
The current-spin correlation function for the case of the Rashba interaction in the linear order, $C^{\alpha\beta}_{js{\rm R}} $, is
(diagrammatically depicted in Fig. \ref{FIGrashba})
\begin{align}
C^{\alpha\beta}_{js{\rm (R)}}(\qv,\pv) &\equiv 
2i\epsilon_{ij\beta} \alpha_{{\rm R},i} (\pv) \biggl[ 
\frac{1}{m}  \lt(k+\frac{q+p}{2}\rt)_\alpha \lt(k+q+\frac{p}{2}\rt)_j  
 2\Re \lt[ g^\ret_{\kv+\qv}g^\ret_{\kv+\qv+\pv} g^\adv_{\kv+\pv}\rt]
 +\delta_{\alpha j} g^\ret_{\kv+\qv} g^\adv_{\kv} \biggr],
\end{align}
where $\qv$ and $\pv$ are wave vectors carried by the external spin source and the Rashba interaction, respectively. 
As is easily checked, the correlation function vanishes for totally uniform case, $\qv=\pv=0$, and at the linear order in $\qv$ or $\pv$.
Expanding with respect to $q$ and $p$ assuming both external field and the Rashba interaction are slowly-varying, we obtain the lowest contribution as 
\begin{align}
C_{js{\rm (R)}}^{\alpha\beta} & 
= \frac{i}{3m^2} \epsilon_{ij\beta} \alpha_{{\rm R},i} (\pv)  
\lt[ -\delta_{\alpha j}(\pv\cdot\qv)+p_\alpha q_j \rt] \sumkv k^2|g^\ret_\kv|^4,
\label{chijsssrashbaq}
\end{align}
which is Eq. (\ref{chijsssrashba}).


\begin{thebibliography}{32}%
\makeatletter
\providecommand \@ifxundefined [1]{%
 \@ifx{#1\undefined}
}%
\providecommand \@ifnum [1]{%
 \ifnum #1\expandafter \@firstoftwo
 \else \expandafter \@secondoftwo
 \fi
}%
\providecommand \@ifx [1]{%
 \ifx #1\expandafter \@firstoftwo
 \else \expandafter \@secondoftwo
 \fi
}%
\providecommand \natexlab [1]{#1}%
\providecommand \enquote  [1]{``#1''}%
\providecommand \bibnamefont  [1]{#1}%
\providecommand \bibfnamefont [1]{#1}%
\providecommand \citenamefont [1]{#1}%
\providecommand \href@noop [0]{\@secondoftwo}%
\providecommand \href [0]{\begingroup \@sanitize@url \@href}%
\providecommand \@href[1]{\@@startlink{#1}\@@href}%
\providecommand \@@href[1]{\endgroup#1\@@endlink}%
\providecommand \@sanitize@url [0]{\catcode `\\12\catcode `\$12\catcode
  `\&12\catcode `\#12\catcode `\^12\catcode `\_12\catcode `\%12\relax}%
\providecommand \@@startlink[1]{}%
\providecommand \@@endlink[0]{}%
\providecommand \url  [0]{\begingroup\@sanitize@url \@url }%
\providecommand \@url [1]{\endgroup\@href {#1}{\urlprefix }}%
\providecommand \urlprefix  [0]{URL }%
\providecommand \Eprint [0]{\href }%
\providecommand \doibase [0]{http://dx.doi.org/}%
\providecommand \selectlanguage [0]{\@gobble}%
\providecommand \bibinfo  [0]{\@secondoftwo}%
\providecommand \bibfield  [0]{\@secondoftwo}%
\providecommand \translation [1]{[#1]}%
\providecommand \BibitemOpen [0]{}%
\providecommand \bibitemStop [0]{}%
\providecommand \bibitemNoStop [0]{.\EOS\space}%
\providecommand \EOS [0]{\spacefactor3000\relax}%
\providecommand \BibitemShut  [1]{\csname bibitem#1\endcsname}%
\let\auto@bib@innerbib\@empty
\bibitem [{\citenamefont {Dyakonov}\ and\ \citenamefont
  {Perel}(1971)}]{Dyakonov71}%
  \BibitemOpen
  \bibfield  {author} {\bibinfo {author} {\bibfnamefont {M.}~\bibnamefont
  {Dyakonov}}\ and\ \bibinfo {author} {\bibfnamefont {V.~I.}\ \bibnamefont
  {Perel}},\ }\href@noop {} {\bibfield  {journal} {\bibinfo  {journal} {Sov.
  Phys. JETP Lett.}\ }\textbf {\bibinfo {volume} {13}},\ \bibinfo {pages} {467}
  (\bibinfo {year} {1971})}\BibitemShut {NoStop}%
\bibitem [{\citenamefont {Hirsch}(1999)}]{Hirsch99}%
  \BibitemOpen
  \bibfield  {author} {\bibinfo {author} {\bibfnamefont {J.~E.}\ \bibnamefont
  {Hirsch}},\ }\href {\doibase 10.1103/PhysRevLett.83.1834} {\bibfield
  {journal} {\bibinfo  {journal} {Phys. Rev. Lett.}\ }\textbf {\bibinfo
  {volume} {83}},\ \bibinfo {pages} {1834} (\bibinfo {year}
  {1999})}\BibitemShut {NoStop}%
\bibitem [{\citenamefont {Kato}\ \emph {et~al.}(2004)\citenamefont {Kato},
  \citenamefont {Myers}, \citenamefont {Gossard},\ and\ \citenamefont
  {Awschalom}}]{Kato04}%
  \BibitemOpen
  \bibfield  {author} {\bibinfo {author} {\bibfnamefont {Y.}~\bibnamefont
  {Kato}}, \bibinfo {author} {\bibfnamefont {R.~C.}\ \bibnamefont {Myers}},
  \bibinfo {author} {\bibfnamefont {A.~C.}\ \bibnamefont {Gossard}}, \ and\
  \bibinfo {author} {\bibfnamefont {D.~D.}\ \bibnamefont {Awschalom}},\
  }\href@noop {} {\bibfield  {journal} {\bibinfo  {journal} {Science}\ }\textbf
  {\bibinfo {volume} {306}},\ \bibinfo {pages} {1910} (\bibinfo {year}
  {2004})}\BibitemShut {NoStop}%
\bibitem [{\citenamefont {Tatara}\ and\ \citenamefont
  {Mizukami}(2017)}]{TataraSP17}%
  \BibitemOpen
  \bibfield  {author} {\bibinfo {author} {\bibfnamefont {G.}~\bibnamefont
  {Tatara}}\ and\ \bibinfo {author} {\bibfnamefont {S.}~\bibnamefont
  {Mizukami}},\ }\href {\doibase 10.1103/PhysRevB.96.064423} {\bibfield
  {journal} {\bibinfo  {journal} {Phys. Rev. B}\ }\textbf {\bibinfo {volume}
  {96}},\ \bibinfo {pages} {064423} (\bibinfo {year} {2017})}\BibitemShut
  {NoStop}%
\bibitem [{\citenamefont {Silsbee}\ \emph {et~al.}(1979)\citenamefont
  {Silsbee}, \citenamefont {Janossy},\ and\ \citenamefont {Monod}}]{Silsbee79}%
  \BibitemOpen
  \bibfield  {author} {\bibinfo {author} {\bibfnamefont {R.~H.}\ \bibnamefont
  {Silsbee}}, \bibinfo {author} {\bibfnamefont {A.}~\bibnamefont {Janossy}}, \
  and\ \bibinfo {author} {\bibfnamefont {P.}~\bibnamefont {Monod}},\ }\href
  {\doibase 10.1103/PhysRevB.19.4382} {\bibfield  {journal} {\bibinfo
  {journal} {Phys. Rev. B}\ }\textbf {\bibinfo {volume} {19}},\ \bibinfo
  {pages} {4382} (\bibinfo {year} {1979})}\BibitemShut {NoStop}%
\bibitem [{\citenamefont {Tserkovnyak}\ \emph {et~al.}(2002)\citenamefont
  {Tserkovnyak}, \citenamefont {Brataas},\ and\ \citenamefont
  {Bauer}}]{Tserkovnyak02}%
  \BibitemOpen
  \bibfield  {author} {\bibinfo {author} {\bibfnamefont {Y.}~\bibnamefont
  {Tserkovnyak}}, \bibinfo {author} {\bibfnamefont {A.}~\bibnamefont
  {Brataas}}, \ and\ \bibinfo {author} {\bibfnamefont {G.~E.~W.}\ \bibnamefont
  {Bauer}},\ }\href {http://link.aps.org/abstract/PRL/v88/e117601} {\bibfield
  {journal} {\bibinfo  {journal} {Phys. Rev. Lett.}\ }\textbf {\bibinfo
  {volume} {88}},\ \bibinfo {eid} {117601} (\bibinfo {year}
  {2002})}\BibitemShut {NoStop}%
\bibitem [{\citenamefont {Burkov}\ \emph {et~al.}(2004)\citenamefont {Burkov},
  \citenamefont {N\'u\~nez},\ and\ \citenamefont {MacDonald}}]{Burkov04}%
  \BibitemOpen
  \bibfield  {author} {\bibinfo {author} {\bibfnamefont {A.~A.}\ \bibnamefont
  {Burkov}}, \bibinfo {author} {\bibfnamefont {A.~S.}\ \bibnamefont
  {N\'u\~nez}}, \ and\ \bibinfo {author} {\bibfnamefont {A.~H.}\ \bibnamefont
  {MacDonald}},\ }\href {\doibase 10.1103/PhysRevB.70.155308} {\bibfield
  {journal} {\bibinfo  {journal} {Phys. Rev. B}\ }\textbf {\bibinfo {volume}
  {70}},\ \bibinfo {pages} {155308} (\bibinfo {year} {2004})}\BibitemShut
  {NoStop}%
\bibitem [{\citenamefont {Galitski}\ \emph {et~al.}(2006)\citenamefont
  {Galitski}, \citenamefont {Burkov},\ and\ \citenamefont
  {Das~Sarma}}]{Galitski06}%
  \BibitemOpen
  \bibfield  {author} {\bibinfo {author} {\bibfnamefont {V.~M.}\ \bibnamefont
  {Galitski}}, \bibinfo {author} {\bibfnamefont {A.~A.}\ \bibnamefont
  {Burkov}}, \ and\ \bibinfo {author} {\bibfnamefont {S.}~\bibnamefont
  {Das~Sarma}},\ }\href {\doibase 10.1103/PhysRevB.74.115331} {\bibfield
  {journal} {\bibinfo  {journal} {Phys. Rev. B}\ }\textbf {\bibinfo {volume}
  {74}},\ \bibinfo {pages} {115331} (\bibinfo {year} {2006})}\BibitemShut
  {NoStop}%
\bibitem [{\citenamefont {Tokatly}\ and\ \citenamefont
  {Sherman}(2010)}]{Tokatly10}%
  \BibitemOpen
  \bibfield  {author} {\bibinfo {author} {\bibfnamefont {I.~V.}\ \bibnamefont
  {Tokatly}}\ and\ \bibinfo {author} {\bibfnamefont {E.~Y.}\ \bibnamefont
  {Sherman}},\ }\href {\doibase 10.1103/PhysRevB.82.161305} {\bibfield
  {journal} {\bibinfo  {journal} {Phys. Rev. B}\ }\textbf {\bibinfo {volume}
  {82}},\ \bibinfo {pages} {161305} (\bibinfo {year} {2010})}\BibitemShut
  {NoStop}%
\bibitem [{\citenamefont {Shibata}\ and\ \citenamefont
  {Kohno}(2011)}]{Shibata11}%
  \BibitemOpen
  \bibfield  {author} {\bibinfo {author} {\bibfnamefont {J.}~\bibnamefont
  {Shibata}}\ and\ \bibinfo {author} {\bibfnamefont {H.}~\bibnamefont
  {Kohno}},\ }\href {\doibase 10.1103/PhysRevB.84.184408} {\bibfield  {journal}
  {\bibinfo  {journal} {Phys. Rev. B}\ }\textbf {\bibinfo {volume} {84}},\
  \bibinfo {pages} {184408} (\bibinfo {year} {2011})}\BibitemShut {NoStop}%
\bibitem [{\citenamefont {Tatara}\ and\ \citenamefont {Kohno}(2004)}]{TK04}%
  \BibitemOpen
  \bibfield  {author} {\bibinfo {author} {\bibfnamefont {G.}~\bibnamefont
  {Tatara}}\ and\ \bibinfo {author} {\bibfnamefont {H.}~\bibnamefont {Kohno}},\
  }\href {http://link.aps.org/abstract/PRL/v92/e086601} {\bibfield  {journal}
  {\bibinfo  {journal} {Phys. Rev. Lett.}\ }\textbf {\bibinfo {volume} {92}},\
  \bibinfo {eid} {086601} (\bibinfo {year} {2004})}\BibitemShut {NoStop}%
\bibitem [{\citenamefont {Tatara}\ \emph {et~al.}(2008)\citenamefont {Tatara},
  \citenamefont {Kohno},\ and\ \citenamefont {Shibata}}]{TKS_PR08}%
  \BibitemOpen
  \bibfield  {author} {\bibinfo {author} {\bibfnamefont {G.}~\bibnamefont
  {Tatara}}, \bibinfo {author} {\bibfnamefont {H.}~\bibnamefont {Kohno}}, \
  and\ \bibinfo {author} {\bibfnamefont {J.}~\bibnamefont {Shibata}},\ }\href
  {\doibase doi:10.1016/j.physrep.2008.07.003} {\bibfield  {journal} {\bibinfo
  {journal} {Physics Reports}\ }\textbf {\bibinfo {volume} {468}},\ \bibinfo
  {pages} {213} (\bibinfo {year} {2008})}\BibitemShut {NoStop}%
\bibitem [{\citenamefont {Yuan}\ and\ \citenamefont {Kelly}(2016)}]{YuanSOT16}%
  \BibitemOpen
  \bibfield  {author} {\bibinfo {author} {\bibfnamefont {Z.}~\bibnamefont
  {Yuan}}\ and\ \bibinfo {author} {\bibfnamefont {P.~J.}\ \bibnamefont
  {Kelly}},\ }\href {\doibase 10.1103/PhysRevB.93.224415} {\bibfield  {journal}
  {\bibinfo  {journal} {Phys. Rev. B}\ }\textbf {\bibinfo {volume} {93}},\
  \bibinfo {pages} {224415} (\bibinfo {year} {2016})}\BibitemShut {NoStop}%
\bibitem [{\citenamefont {Ado}\ \emph {et~al.}(2017)\citenamefont {Ado},
  \citenamefont {Tretiakov},\ and\ \citenamefont {Titov}}]{Ado17}%
  \BibitemOpen
  \bibfield  {author} {\bibinfo {author} {\bibfnamefont {I.~A.}\ \bibnamefont
  {Ado}}, \bibinfo {author} {\bibfnamefont {O.~A.}\ \bibnamefont {Tretiakov}},
  \ and\ \bibinfo {author} {\bibfnamefont {M.}~\bibnamefont {Titov}},\ }\href
  {\doibase 10.1103/PhysRevB.95.094401} {\bibfield  {journal} {\bibinfo
  {journal} {Phys. Rev. B}\ }\textbf {\bibinfo {volume} {95}},\ \bibinfo
  {pages} {094401} (\bibinfo {year} {2017})}\BibitemShut {NoStop}%
\bibitem [{\citenamefont {Cr\'epieux}\ and\ \citenamefont
  {Bruno}(2001)}]{Crepieux01}%
  \BibitemOpen
  \bibfield  {author} {\bibinfo {author} {\bibfnamefont {A.}~\bibnamefont
  {Cr\'epieux}}\ and\ \bibinfo {author} {\bibfnamefont {P.}~\bibnamefont
  {Bruno}},\ }\href {\doibase 10.1103/PhysRevB.64.014416} {\bibfield  {journal}
  {\bibinfo  {journal} {Phys. Rev. B}\ }\textbf {\bibinfo {volume} {64}},\
  \bibinfo {pages} {014416} (\bibinfo {year} {2001})}\BibitemShut {NoStop}%
\bibitem [{\citenamefont {Saitoh}\ \emph {et~al.}(2006)\citenamefont {Saitoh},
  \citenamefont {Ueda}, \citenamefont {Miyajima},\ and\ \citenamefont
  {Tatara}}]{Saitoh06}%
  \BibitemOpen
  \bibfield  {author} {\bibinfo {author} {\bibfnamefont {E.}~\bibnamefont
  {Saitoh}}, \bibinfo {author} {\bibfnamefont {M.}~\bibnamefont {Ueda}},
  \bibinfo {author} {\bibfnamefont {H.}~\bibnamefont {Miyajima}}, \ and\
  \bibinfo {author} {\bibfnamefont {G.}~\bibnamefont {Tatara}},\ }\href
  {\doibase 10.1063/1.2199473} {\bibfield  {journal} {\bibinfo  {journal}
  {Applied Physics Letters}\ }\textbf {\bibinfo {volume} {88}},\ \bibinfo
  {pages} {182509} (\bibinfo {year} {2006})},\ \Eprint
  {http://arxiv.org/abs/http://dx.doi.org/10.1063/1.2199473}
  {http://dx.doi.org/10.1063/1.2199473} \BibitemShut {NoStop}%
\bibitem [{\citenamefont {Tatara}\ and\ \citenamefont {Entel}(2008)}]{TE08}%
  \BibitemOpen
  \bibfield  {author} {\bibinfo {author} {\bibfnamefont {G.}~\bibnamefont
  {Tatara}}\ and\ \bibinfo {author} {\bibfnamefont {P.}~\bibnamefont {Entel}},\
  }\href {\doibase 10.1103/PhysRevB.78.064429} {\bibfield  {journal} {\bibinfo
  {journal} {Phys. Rev. B}\ }\textbf {\bibinfo {volume} {78}},\ \bibinfo {eid}
  {064429} (\bibinfo {year} {2008})}\BibitemShut {NoStop}%
\bibitem [{\citenamefont {Nakabayashi}\ \emph {et~al.}(2010)\citenamefont
  {Nakabayashi}, \citenamefont {Takeuchi}, \citenamefont {Hosono},
  \citenamefont {Taguchi},\ and\ \citenamefont {Tatara}}]{Nakabayashi10}%
  \BibitemOpen
  \bibfield  {author} {\bibinfo {author} {\bibfnamefont {N.}~\bibnamefont
  {Nakabayashi}}, \bibinfo {author} {\bibfnamefont {A.}~\bibnamefont
  {Takeuchi}}, \bibinfo {author} {\bibfnamefont {K.}~\bibnamefont {Hosono}},
  \bibinfo {author} {\bibfnamefont {K.}~\bibnamefont {Taguchi}}, \ and\
  \bibinfo {author} {\bibfnamefont {G.}~\bibnamefont {Tatara}},\ }\href
  {\doibase 10.1103/PhysRevB.82.014403} {\bibfield  {journal} {\bibinfo
  {journal} {Phys. Rev. B}\ }\textbf {\bibinfo {volume} {82}},\ \bibinfo
  {pages} {014403} (\bibinfo {year} {2010})}\BibitemShut {NoStop}%
\bibitem [{\citenamefont {Hua~Chen}(2018)}]{Chen18}%
  \BibitemOpen
  \bibfield  {author} {\bibinfo {author} {\bibfnamefont {A.~H.~M.}\
  \bibnamefont {Hua~Chen}, \bibfnamefont {Qian~Niu}},\ }\href@noop {}
  {\bibfield  {journal} {\bibinfo  {journal} {arXiv:1803.01294}\ } (\bibinfo
  {year} {2018})}\BibitemShut {NoStop}%
\bibitem [{\citenamefont {Kawaguchi}\ and\ \citenamefont
  {Tatara}(2016)}]{Kawaguchi16}%
  \BibitemOpen
  \bibfield  {author} {\bibinfo {author} {\bibfnamefont {H.}~\bibnamefont
  {Kawaguchi}}\ and\ \bibinfo {author} {\bibfnamefont {G.}~\bibnamefont
  {Tatara}},\ }\href {\doibase 10.1103/PhysRevB.94.235148} {\bibfield
  {journal} {\bibinfo  {journal} {Phys. Rev. B}\ }\textbf {\bibinfo {volume}
  {94}},\ \bibinfo {pages} {235148} (\bibinfo {year} {2016})}\BibitemShut
  {NoStop}%
\bibitem [{\citenamefont {Kawaguchi}\ and\ \citenamefont
  {Tatara}(2018)}]{Kawaguchi18}%
  \BibitemOpen
  \bibfield  {author} {\bibinfo {author} {\bibfnamefont {H.}~\bibnamefont
  {Kawaguchi}}\ and\ \bibinfo {author} {\bibfnamefont {G.}~\bibnamefont
  {Tatara}},\ }\href {\doibase 10.7566/JPSJ.87.064002} {\bibfield  {journal}
  {\bibinfo  {journal} {Journal of the Physical Society of Japan}\ }\textbf
  {\bibinfo {volume} {87}},\ \bibinfo {pages} {064002} (\bibinfo {year}
  {2018})},\ \Eprint
  {http://arxiv.org/abs/https://journals.jps.jp/doi/pdf/10.7566/JPSJ.87.064002}
  {https://journals.jps.jp/doi/pdf/10.7566/JPSJ.87.064002} \BibitemShut
  {NoStop}%
\bibitem [{\citenamefont {Sanchez}\ \emph {et~al.}(2013)\citenamefont
  {Sanchez}, \citenamefont {Vila}, \citenamefont {Desfonds}, \citenamefont
  {Gambarelli}, \citenamefont {Attane}, \citenamefont {De~Teresa},
  \citenamefont {Magen},\ and\ \citenamefont {Fert}}]{Sanchez13}%
  \BibitemOpen
  \bibfield  {author} {\bibinfo {author} {\bibfnamefont {J.~C.~R.}\
  \bibnamefont {Sanchez}}, \bibinfo {author} {\bibfnamefont {L.}~\bibnamefont
  {Vila}}, \bibinfo {author} {\bibfnamefont {G.}~\bibnamefont {Desfonds}},
  \bibinfo {author} {\bibfnamefont {S.}~\bibnamefont {Gambarelli}}, \bibinfo
  {author} {\bibfnamefont {J.~P.}\ \bibnamefont {Attane}}, \bibinfo {author}
  {\bibfnamefont {J.~M.}\ \bibnamefont {De~Teresa}}, \bibinfo {author}
  {\bibfnamefont {C.}~\bibnamefont {Magen}}, \ and\ \bibinfo {author}
  {\bibfnamefont {A.}~\bibnamefont {Fert}},\ }\href
  {http://dx.doi.org/10.1038/ncomms3944} {\bibfield  {journal} {\bibinfo
  {journal} {Nat Commun}\ }\textbf {\bibinfo {volume} {4}},\ \bibinfo {pages}
  {2944} (\bibinfo {year} {2013})},\ \bibinfo {note} {article}\BibitemShut
  {NoStop}%
\bibitem [{\citenamefont {Edelstein}(1990)}]{Edelstein90}%
  \BibitemOpen
  \bibfield  {author} {\bibinfo {author} {\bibfnamefont {V.}~\bibnamefont
  {Edelstein}},\ }\href {\doibase 10.1016/0038-1098(90)90963-C} {\bibfield
  {journal} {\bibinfo  {journal} {Solid State Communications}\ }\textbf
  {\bibinfo {volume} {73}},\ \bibinfo {pages} {233 } (\bibinfo {year}
  {1990})}\BibitemShut {NoStop}%
\bibitem [{\citenamefont {Shen}\ \emph {et~al.}(2014)\citenamefont {Shen},
  \citenamefont {Vignale},\ and\ \citenamefont {Raimondi}}]{Shen14}%
  \BibitemOpen
  \bibfield  {author} {\bibinfo {author} {\bibfnamefont {K.}~\bibnamefont
  {Shen}}, \bibinfo {author} {\bibfnamefont {G.}~\bibnamefont {Vignale}}, \
  and\ \bibinfo {author} {\bibfnamefont {R.}~\bibnamefont {Raimondi}},\ }\href
  {\doibase 10.1103/PhysRevLett.112.096601} {\bibfield  {journal} {\bibinfo
  {journal} {Phys. Rev. Lett.}\ }\textbf {\bibinfo {volume} {112}},\ \bibinfo
  {pages} {096601} (\bibinfo {year} {2014})}\BibitemShut {NoStop}%
\bibitem [{\citenamefont {Kim}\ \emph {et~al.}(2012)\citenamefont {Kim},
  \citenamefont {Moon}, \citenamefont {Lee},\ and\ \citenamefont
  {Lee}}]{Kim12}%
  \BibitemOpen
  \bibfield  {author} {\bibinfo {author} {\bibfnamefont {K.-W.}\ \bibnamefont
  {Kim}}, \bibinfo {author} {\bibfnamefont {J.-H.}\ \bibnamefont {Moon}},
  \bibinfo {author} {\bibfnamefont {K.-J.}\ \bibnamefont {Lee}}, \ and\
  \bibinfo {author} {\bibfnamefont {H.-W.}\ \bibnamefont {Lee}},\ }\href
  {\doibase 10.1103/PhysRevLett.108.217202} {\bibfield  {journal} {\bibinfo
  {journal} {Phys. Rev. Lett.}\ }\textbf {\bibinfo {volume} {108}},\ \bibinfo
  {pages} {217202} (\bibinfo {year} {2012})}\BibitemShut {NoStop}%
\bibitem [{\citenamefont {Tatara}\ \emph {et~al.}(2013)\citenamefont {Tatara},
  \citenamefont {Nakabayashi},\ and\ \citenamefont {Lee}}]{Tatara_smf13}%
  \BibitemOpen
  \bibfield  {author} {\bibinfo {author} {\bibfnamefont {G.}~\bibnamefont
  {Tatara}}, \bibinfo {author} {\bibfnamefont {N.}~\bibnamefont {Nakabayashi}},
  \ and\ \bibinfo {author} {\bibfnamefont {K.-J.}\ \bibnamefont {Lee}},\ }\href
  {\doibase 10.1103/PhysRevB.87.054403} {\bibfield  {journal} {\bibinfo
  {journal} {Phys. Rev. B}\ }\textbf {\bibinfo {volume} {87}},\ \bibinfo
  {pages} {054403} (\bibinfo {year} {2013})}\BibitemShut {NoStop}%
\bibitem [{\citenamefont {Nakabayashi}\ and\ \citenamefont
  {Tatara}(2014)}]{Nakabayashi14}%
  \BibitemOpen
  \bibfield  {author} {\bibinfo {author} {\bibfnamefont {N.}~\bibnamefont
  {Nakabayashi}}\ and\ \bibinfo {author} {\bibfnamefont {G.}~\bibnamefont
  {Tatara}},\ }\href {http://stacks.iop.org/1367-2630/16/i=1/a=015016}
  {\bibfield  {journal} {\bibinfo  {journal} {New Journal of Physics}\ }\textbf
  {\bibinfo {volume} {16}},\ \bibinfo {pages} {015016} (\bibinfo {year}
  {2014})}\BibitemShut {NoStop}%
\bibitem [{\citenamefont {Nakazawa}\ and\ \citenamefont
  {Kohno}(2014)}]{Nakazawa14}%
  \BibitemOpen
  \bibfield  {author} {\bibinfo {author} {\bibfnamefont {K.}~\bibnamefont
  {Nakazawa}}\ and\ \bibinfo {author} {\bibfnamefont {H.}~\bibnamefont
  {Kohno}},\ }\href {\doibase 10.7566/JPSJ.83.073707} {\bibfield  {journal}
  {\bibinfo  {journal} {Journal of the Physical Society of Japan}\ }\textbf
  {\bibinfo {volume} {83}},\ \bibinfo {pages} {073707} (\bibinfo {year}
  {2014})},\ \Eprint
  {http://arxiv.org/abs/http://dx.doi.org/10.7566/JPSJ.83.073707}
  {http://dx.doi.org/10.7566/JPSJ.83.073707} \BibitemShut {NoStop}%
\bibitem [{\citenamefont {Qiu}\ \emph {et~al.}(2016)\citenamefont {Qiu},
  \citenamefont {Li}, \citenamefont {Hou}, \citenamefont {Arenholz},
  \citenamefont {N{\^a}??Diaye}, \citenamefont {Tan}, \citenamefont {Uchida},
  \citenamefont {Sato}, \citenamefont {Okamoto}, \citenamefont {Tserkovnyak},
  \citenamefont {Qiu},\ and\ \citenamefont {Saitoh}}]{Qiu16}%
  \BibitemOpen
  \bibfield  {author} {\bibinfo {author} {\bibfnamefont {Z.}~\bibnamefont
  {Qiu}}, \bibinfo {author} {\bibfnamefont {J.}~\bibnamefont {Li}}, \bibinfo
  {author} {\bibfnamefont {D.}~\bibnamefont {Hou}}, \bibinfo {author}
  {\bibfnamefont {E.}~\bibnamefont {Arenholz}}, \bibinfo {author}
  {\bibfnamefont {A.~T.}\ \bibnamefont {N{\^a}??Diaye}}, \bibinfo {author}
  {\bibfnamefont {A.}~\bibnamefont {Tan}}, \bibinfo {author} {\bibfnamefont
  {K.-i.}\ \bibnamefont {Uchida}}, \bibinfo {author} {\bibfnamefont
  {K.}~\bibnamefont {Sato}}, \bibinfo {author} {\bibfnamefont {S.}~\bibnamefont
  {Okamoto}}, \bibinfo {author} {\bibfnamefont {Y.}~\bibnamefont
  {Tserkovnyak}}, \bibinfo {author} {\bibfnamefont {Z.~Q.}\ \bibnamefont
  {Qiu}}, \ and\ \bibinfo {author} {\bibfnamefont {E.}~\bibnamefont {Saitoh}},\
  }\href {http://dx.doi.org/10.1038/ncomms12670} {\bibfield  {journal}
  {\bibinfo  {journal} {Nature Communications}\ }\textbf {\bibinfo {volume}
  {7}},\ \bibinfo {pages} {12670 EP } (\bibinfo {year} {2016})},\ \bibinfo
  {note} {article}\BibitemShut {NoStop}%
\bibitem [{\citenamefont {Qu}\ \emph {et~al.}(2015)\citenamefont {Qu},
  \citenamefont {Huang},\ and\ \citenamefont {Chien}}]{Qu15}%
  \BibitemOpen
  \bibfield  {author} {\bibinfo {author} {\bibfnamefont {D.}~\bibnamefont
  {Qu}}, \bibinfo {author} {\bibfnamefont {S.~Y.}\ \bibnamefont {Huang}}, \
  and\ \bibinfo {author} {\bibfnamefont {C.~L.}\ \bibnamefont {Chien}},\ }\href
  {\doibase 10.1103/PhysRevB.92.020418} {\bibfield  {journal} {\bibinfo
  {journal} {Phys. Rev. B}\ }\textbf {\bibinfo {volume} {92}},\ \bibinfo
  {pages} {020418} (\bibinfo {year} {2015})}\BibitemShut {NoStop}%
\bibitem [{\citenamefont {Cramer}\ \emph {et~al.}(2018)\citenamefont {Cramer},
  \citenamefont {Ritzmann}, \citenamefont {Dong}, \citenamefont {Jaiswal},
  \citenamefont {Qiu}, \citenamefont {Saitoh}, \citenamefont {Nowak},\ and\
  \citenamefont {Kläui}}]{Cramer18}%
  \BibitemOpen
  \bibfield  {author} {\bibinfo {author} {\bibfnamefont {J.}~\bibnamefont
  {Cramer}}, \bibinfo {author} {\bibfnamefont {U.}~\bibnamefont {Ritzmann}},
  \bibinfo {author} {\bibfnamefont {B.-W.}\ \bibnamefont {Dong}}, \bibinfo
  {author} {\bibfnamefont {S.}~\bibnamefont {Jaiswal}}, \bibinfo {author}
  {\bibfnamefont {Z.}~\bibnamefont {Qiu}}, \bibinfo {author} {\bibfnamefont
  {E.}~\bibnamefont {Saitoh}}, \bibinfo {author} {\bibfnamefont
  {U.}~\bibnamefont {Nowak}}, \ and\ \bibinfo {author} {\bibfnamefont
  {M.}~\bibnamefont {Kläui}},\ }\href
  {http://stacks.iop.org/0022-3727/51/i=14/a=144004} {\bibfield  {journal}
  {\bibinfo  {journal} {Journal of Physics D: Applied Physics}\ }\textbf
  {\bibinfo {volume} {51}},\ \bibinfo {pages} {144004} (\bibinfo {year}
  {2018})}\BibitemShut {NoStop}%
\bibitem [{\citenamefont {Ohnuma}\ \emph {et~al.}(2014)\citenamefont {Ohnuma},
  \citenamefont {Adachi}, \citenamefont {Saitoh},\ and\ \citenamefont
  {Maekawa}}]{Ohnuma14}%
  \BibitemOpen
  \bibfield  {author} {\bibinfo {author} {\bibfnamefont {Y.}~\bibnamefont
  {Ohnuma}}, \bibinfo {author} {\bibfnamefont {H.}~\bibnamefont {Adachi}},
  \bibinfo {author} {\bibfnamefont {E.}~\bibnamefont {Saitoh}}, \ and\ \bibinfo
  {author} {\bibfnamefont {S.}~\bibnamefont {Maekawa}},\ }\href {\doibase
  10.1103/PhysRevB.89.174417} {\bibfield  {journal} {\bibinfo  {journal} {Phys.
  Rev. B}\ }\textbf {\bibinfo {volume} {89}},\ \bibinfo {pages} {174417}
  (\bibinfo {year} {2014})}\BibitemShut {NoStop}%
\end{thebibliography}
%

\end{document}